# Phase-field modeling of elastically driven abnormal grain growth

Yazhuo Liu[a], Yin Zhang[b], Kunqing Ding[a], Yichen Yang[a], Alejandro Barrios[c], Xavier Maeder[d], Olivier Pierron[a], Xing Liu[e,*], Ting Zhu[a,*]

[a]Woodruff School of Mechanical Engineering, Georgia Institute of Technology, Atlanta, GA, 30332, USA

[b]School of Mechanics and Engineering Science, Peking University, 100871, Beijing, China

[c]Colorado School of Mines, Golden, CO, 80401, USA

[d]EMPA, Swiss Federal Laboratories for Materials Science and Technology, Laboratory for Mechanics of Materials and Nanostructures, Thun, 3602, Switzerland

[e]Department of Mechanical and Industrial Engineering, New Jersey Institute of Technology, Newark, NJ, 07102, USA

## Abstract

Grain-refined metals offer high strength, yet their engineering applications are often constrained by grain coarsening under mechanical loading. Recent experiments have revealed room-temperature abnormal grain growth (AGG) in ultrafine-grained Ni thin films subjected to cyclic loading. This phenomenon occurs in the regime of macroscopic elastic deformation, unlike conventional AGG, which generally requires substantial plastic deformation or high temperatures. The AGG process is characterized by the preferential growth of grains whose in-plane ⟨100⟩ orientations are aligned with the loading direction. Here, we investigate the underlying physical mechanisms by combining phase-field simulations with micromechanical analysis. The results indicate that a reduction in elastic energy provides a thermodynamic driving force for orientation-selective grain growth. Phase-field simulations reveal the kinetic evolution of AGG and demonstrate that local grain geometry and stress state influence grain growth pathways. By applying this framework to materials with varying degrees of elastic anisotropy, we establish a general approach for investigating elastically driven AGG in polycrystalline systems.

Corresponding authors: xing.liu@njit.edu (X.L.), ting.zhu@me.gatech.edu (T.Z.)

## 1. Introduction

Abnormal grain growth (AGG) is an orientation-selective mode of grain coarsening in polycrystalline metals (Humphreys et al., 2017) that significantly affects the stability and mechanical properties of fine-grained materials. Unlike normal grain growth, in which grains coarsen uniformly, AGG involves the preferential growth of specific grains at the expense of the surrounding matrix. This selectivity is especially important for ultrafine-grained metals, as their high strength relies on maintaining a refined grain structure. Recent experiments on ultrafine-grained Ni thin films have revealed that AGG can occur during room-temperature cyclic loading within the macroscopic elastic regime (Barrios et al., 2022). This observation is notable because it demonstrates that significant grain coarsening can take place even in the absence of high temperatures or substantial macroscopic plastic deformation.

The mechanisms underlying AGG during macroscopic elastic loading remain unclear. Existing explanations for AGG primarily invoke large plastic deformation (Bednarczyk et al., 2021; Ciulik and Taleff, 2009; Sonnweber-Ribic et al., 2012; Wang et al., 2019), thermal stability (Greiser et al., 1999; Lee et al., 2000; Omori et al., 2013; Thuvander et al., 2001), surface energy anisotropy (Wong et al., 1986), or grain boundary (GB) energy anisotropy (Grest et al., 1990; Peng et al., 2023; Rollett et al., 1989). By comparison, the theoretical understanding of AGG is less developed than that of normal grain growth (Anderson et al., 1989; Kurtz and Carpay, 1980; Louat, 1974). These explanations cannot readily account for room-temperature cyclic loading experiments in which strong orientation selectivity is observed while the macroscopic strain remains within the elastic range (Barrios et al., 2022). Although local inelastic processes at GBs may contribute to this phenomenon, GB microplasticity alone cannot account for the observed orientation selectivity. Consequently, a central question is whether elastic energy, mediated by crystallographic elastic anisotropy, can provide a thermodynamically viable driving force for AGG.

In this study, we combine phase-field simulations (Chen, 2002; Tonks and Millett, 2011; Tonks et al., 2010) with micromechanical analysis (Eshelby, 1957; Qu and Cherkaoui, 2006; Zhang et al., 2020) to investigate AGG under small-strain loading at room temperature. The phase-field model incorporates orientation-dependent elastic energy as the thermodynamic driving force for GB migration, enabling the simulated microstructure to capture the orientation-selective growth behavior in face-centered cubic (FCC) Ni thin films. Micromechanical analysis interprets the simulated GB motion by considering local grain geometry, stress state, and the jump in the Eshelby elastic energy-momentum tensor across GBs. This study reveals how a reduction in elastic energy drives AGG, elucidates why individual GBs may migrate along distinct local pathways, and examines the influence of cyclic loading on time-averaged driving forces and GB mobility. Finally, by extending the analysis to body-centered cubic (BCC) W and Cr with varying degrees of elastic anisotropy, we establish a general approach for studying elastically driven AGG in polycrystalline materials.

## 2. Experimental

AGG has recently been observed in ultrafine-grained FCC Ni microbeams subjected to cyclic bending at room temperature (Barrios et al., 2022). Figure 1 shows a representative example of AGG observed at the bottom of an electroplated Ni microbeam after it underwent 160 million cycles of symmetric cyclic loading with a strain amplitude of 0.26% and at a frequency of 8 kHz (see Supplementary Section 1.1 for the experimental setup). The as-deposited microstructure exhibited a strong out-of-plane [001] texture (Fig. 1(a)) with random in-plane grain orientations (Fig. 1(b)). The area-weighted average grain size is less than ~640 nm within the bottom ~2 μm of the microbeam. After cyclic loading, a dominant in-plane texture developed in the near-surface regions (i.e., at the edges of the microbeam) subjected to peak tensile and compressive stresses, whereas the out-of-plane texture remained unchanged. Specifically, grains with an in-plane [100] orientation along the loading direction underwent preferential lateral growth (Fig. 1(c)).

The estimated applied stress in the near-surface regions was ~0.6 GPa (Zacharias, 1933). This stress is below the predicted yield strength (~1.1 GPa) and thus within the macroscopic elastic regime (Zhou et al., 2020), thereby limiting significant macroscopic plastic deformation (Barrios et al., 2022). Based on post-mortem observations of the grain structure, the estimated effective GB mobility was on the order of $10^{-3}$ μm $s^{-1}$ $MPa^{-1}$. These results indicate that substantial grain growth can be induced by low-amplitude cyclic loading at room temperature.

Notably, no significant grain growth was observed when a constant (static) load was applied for a duration comparable to that used in the cyclic-loading experiments. This finding suggests that cyclic loading facilitates GB migration associated with AGG. Specifically, repeated stress reversals may promote local microplasticity at GBs, whereas such microplasticity is absent or substantially reduced under static loading conditions. To further clarify the effects of cyclic loading, additional low-frequency (0.5 Hz) experiments were conducted. Despite a reduction in loading frequency of several orders of magnitude, grains with their in-plane [100] direction aligned with the loading direction continued to exhibit preferential growth. However, the effective GB mobility decreased by several orders of magnitude, e.g., from $10^{-2}$ to $10^{-5}$ μm $s^{-1}$ $MPa^{-1}$ (Supplementary Section 1.2). These observations support the proposed kinetic role of cyclic loading in promoting GB migration and, consequently, AGG. Similar orientation-selective growth was observed in parallel experiments on randomly textured Ni microbeams, in which prolonged cyclic loading produced an exceptionally large grain with its [100] direction aligned with the loading direction.

Although the applied stress was below the macroscopic yield strength, the possibility of local inelastic processes occurring at GBs cannot be ruled out. Driven by stress non-uniformity and intergranular interactions, local stresses at GBs may exceed the macroscopic stress level, thereby activating localized microplasticity without inducing bulk yielding. However, while these local processes may facilitate GB migration, they do not inherently generate crystallographic selectivity. Consequently, the consistent preferential growth of a specific grain family across varying loading frequencies or amplitudes suggests the existence of an orientation-dependent elastic driving force.

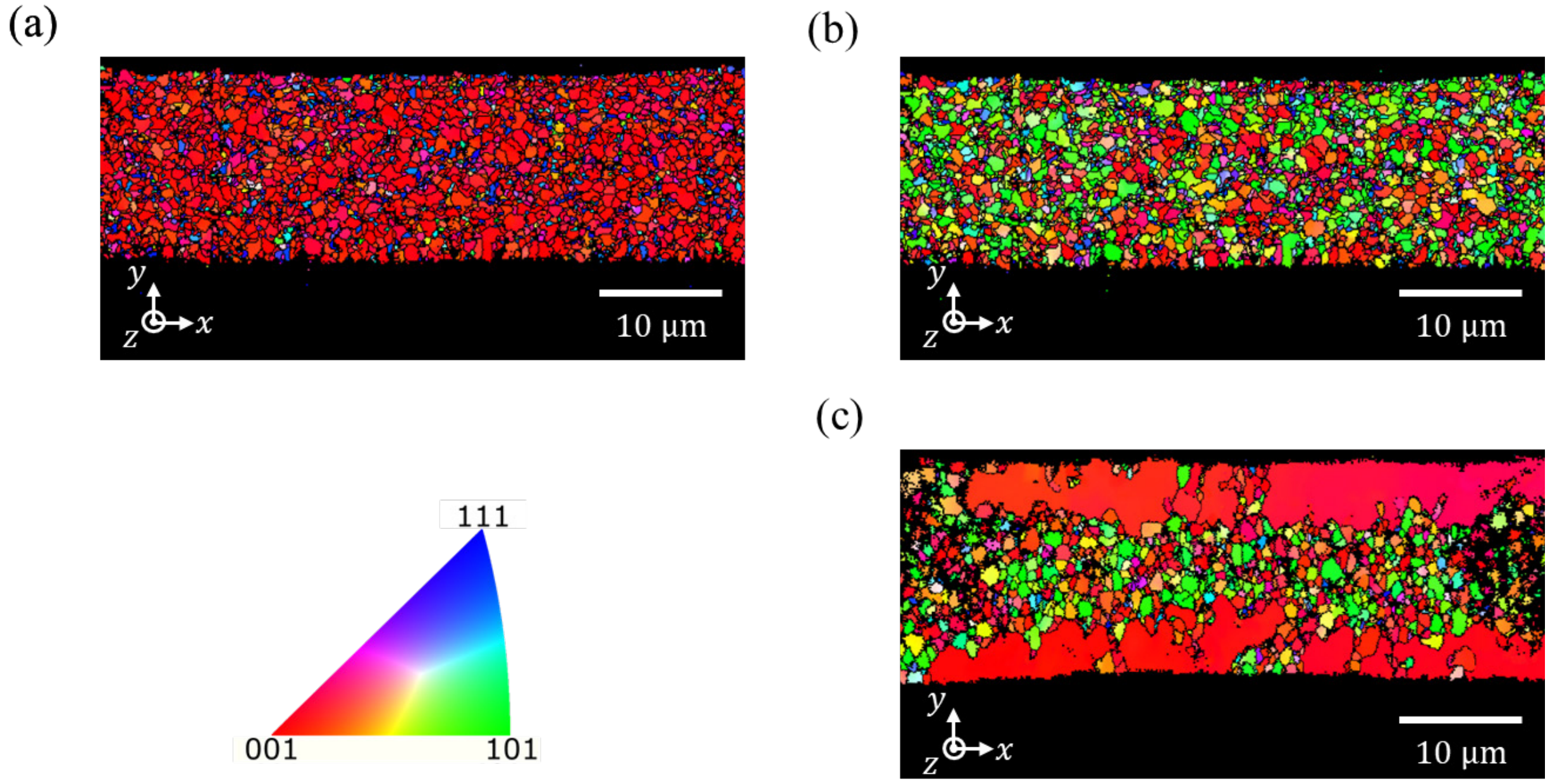

**Fig. 1.** Experimental observation of AGG in an ultrafine-grained Ni microbeam under small-strain cyclic loading at room temperature. (a) Initial out-of-plane grain orientations (*z*-axis), showing a predominant [001] texture. (b-c) In-plane grain orientations along the loading direction (*x*-axis) (b) before and (c) after $1.6 \times 10^8$ loading cycles, respectively. Grains from the near-[100] orientation

family (red) grow preferentially, consuming those of other orientations.

## 3. General considerations

Several mechanisms may explain the [100]-predominant grain growth. Previous work demonstrated that differences in GB mobility arising from GB complexion transitions can produce AGG (Marvel et al., 2022). In the present system, however, attributing AGG solely to mobility differences would require a mechanism that systematically links enhanced mobility to a particular in-plane crystallographic direction. We have found no evidence of such a link. Because the in-plane grain orientations are largely random in our samples, the GBs surrounding grains whose [100] direction is along the loading direction are not expected to form a distinct population characterized by systematically lower energy, higher mobility, or a greater propensity for defect-induced structural transformation.

In principle, crystallographic orientations that favor cyclic slip could also promote preferential growth. Local slip processes may be essential for activating GB migration under cyclic loading, especially because no appreciable AGG occurs under comparable static loading conditions. However, this mechanism cannot explain the preferential growth of grains with [100] aligned with the loading direction, because this alignment is unfavorable for slip activity. Micromechanical analysis indicates that, in FCC polycrystals, the [100] grain family exhibits the lowest maximum resolved shear stress normalized by the applied stress (Chen et al., 2019). Thus, the grain family that grows preferentially in our experiments is the least favorably oriented for cyclic slip, indicating that a different driving force governs this behavior.

Another possible mechanism for the observed AGG phenomenon is a reduction in elastic energy. For a polycrystalline sample of volume $V$ subjected to stress-controlled loading in the elastic regime, the total potential energy $\Pi$, a standard quantity in linear elasticity, should be considered. Under an applied uniaxial stress $\sigma_0$, $\Pi$ at static equilibrium is the sum of the strain energy, $\frac{\sigma_0^2}{2\bar{E}}V$, and the work performed by the external loads, $-\frac{\sigma_0^2}{\bar{E}}V$, yielding

$$\Pi = -\left(\frac{\sigma_0^2}{2\bar{E}}\right)V \tag{1}$$

where $\bar{E}$ is the effective Young's modulus of the polycrystal along the loading direction. Eq. (1) indicates that AGG reduces $\Pi$ when elastically soft grains expand at the expense of stiffer ones, since this lowers $\bar{E}$ and, consequently, $\Pi$. A similar argument applies under strain-controlled loading, where the elastic strain energy governs the driving force: the growth of elastically soft grains lowers the stored elastic energy of the polycrystal. Although the specific energy terms differ between the stress- and strain-controlled conditions, both cases reflect the same principle: AGG lowers elastic deformation energy. Hence, we attribute the driving force for AGG to a reduction in elastic energy. Since $\Pi$ depends quadratically on $\sigma_0$, its reduction is independent of the sign of $\sigma_0$. Consequently, under symmetric cyclic loading, the cycle-averaged energetic driving force can be represented by an effective time-averaged elastic driving force comparable to that generated by a static stress with magnitude $\sigma_0$ (see Section 5.4 for a more detailed discussion). This treatment does not imply that cyclic and monotonic loading are identical. Rather, while both yield comparable thermodynamic driving forces, cyclic loading may exert a distinct kinetic influence by accelerating GB migration through localized microplasticity and defect interactions. Under these conditions, elastic anisotropy remains the thermodynamic driver governing the preferential growth of elastically compliant grain orientations.

For a polycrystal with random grain orientations, the key question is which orientation has the lowest elastic stiffness under the applied loading direction. To address this question, we consider a representative spherical grain of cubic symmetry embedded in a polycrystal subjected to uniaxial tension $\sigma_0$ (Fig. 2(a)). The stress and strain in this grain can be evaluated using a micromechanics analysis that combines the classical Eshelby inclusion solution (Eshelby, 1957) with the self-consistent solution for the effective elastic moduli of polycrystals (Qu and Cherkaoui, 2006; Zhang et al., 2020). Although individual grains are elastically anisotropic, the random distribution of grain orientations leads to an effectively uniform, isotropic elastic modulus at the polycrystal level. According to Eshelby's solution (Eshelby, 1957), the stress, strain, and strain energy density within a spherical grain embedded in a homogenized polycrystal (Fig. 2(b)) are uniform. Suppose this grain belongs to a $\{hkl\}$ family, with the corresponding $\langle hkl\rangle$ direction aligned with the loading axis. As the polycrystal is subjected to $\sigma_0$, the average tensile strain in the $\{hkl\}$ family is

$$\varepsilon^{hkl} = \frac{\sigma_0}{E^{hkl}} \tag{2}$$

where $E^{hkl}$ is the effective elastic stiffness along the loading direction (also referred to as the diffraction elastic constant) and can be derived as follows (Zhang et al., 2020)

$$\frac{1}{E^{hkl}} = \frac{3a + 4b}{3} - 4(b - c)\Gamma \quad (3)$$

with $a$, $b$, and $c$ being parameters determined by the single-crystal elastic constants (see (Zhang et al., 2020) for detailed expressions), and $\Gamma$ being the orientation index defined as

$$\Gamma = \frac{h^2k^2 + l^2k^2 + h^2l^2}{(h^2 + k^2 + l^2)^2} \quad (4)$$

For an FCC Ni polycrystal with randomly oriented grains, whose single-crystal elastic constants are listed in Table 1, Fig. 2(c) shows the relation between $E^{hkl}$ and $\Gamma$ based on Eq. (3). The orientation index $\Gamma$ ranges from 0 to 1/3, with representative values of 0, 19/121, 1/4 and 1/3 for the $\{100\}$, $\{311\}$, $\{110\}$ and $\{111\}$ families, respectively. Since $E^{100}$ is the lowest, the $\{100\}$ family is the elastically softest along the loading direction. Hence, AGG leads to preferential alignment of in-plane $\langle 100 \rangle$ orientations along the loading direction.

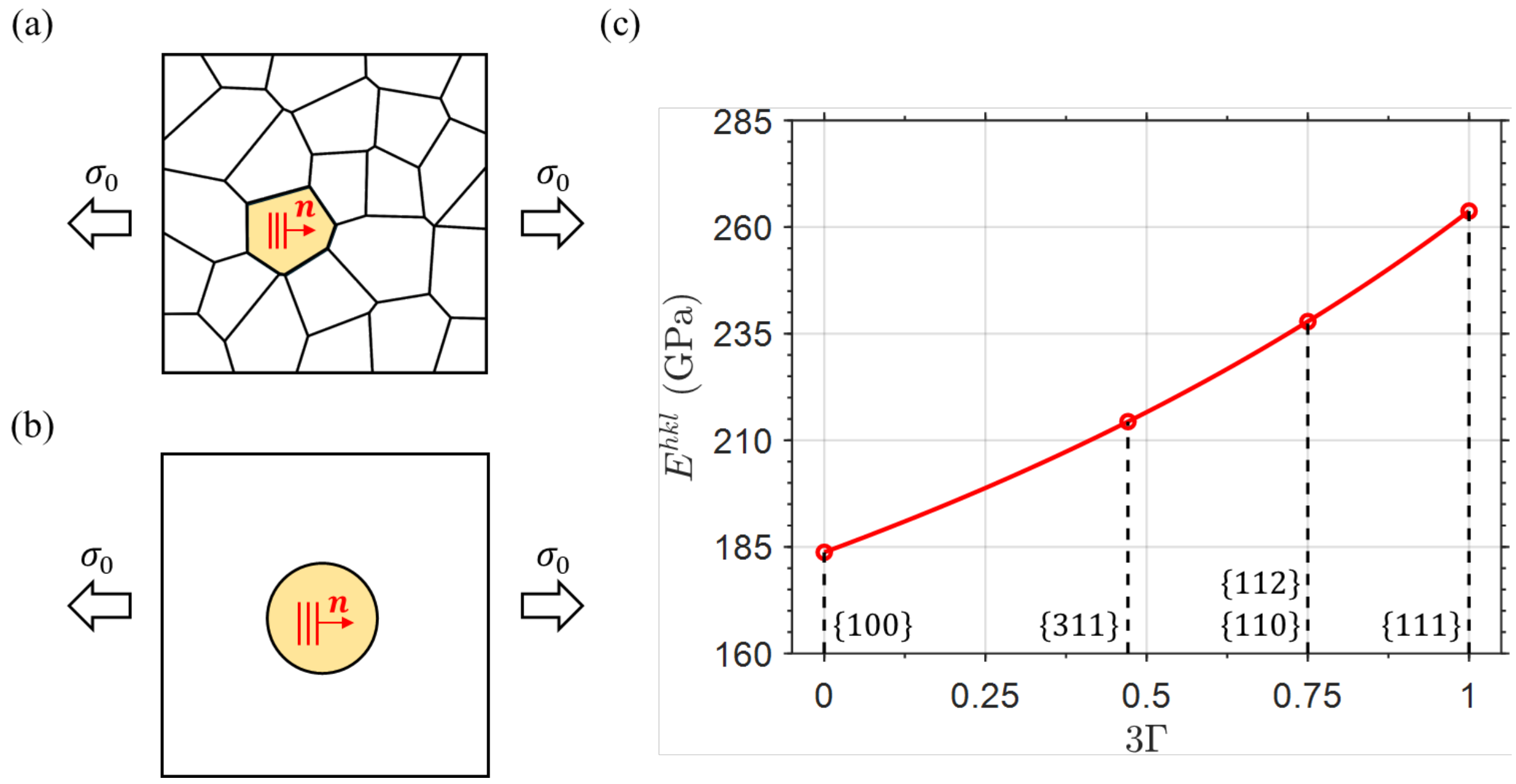

**Fig. 2**. Micromechanical analysis of grain orientation effects. (a) In a polycrystal subjected to uniaxial tension $\sigma_0$, a representative grain with orientation $\boldsymbol{n}$ along the loading direction is highlighted in yellow. (b) The yellow grain in (a) is modeled as a circular grain embedded in a homogenized polycrystal. (c) Effective elastic stiffness along the loading direction for the circular grain in (b) as a function of orientation index $\Gamma$.

The AGG phenomenon described in Section 2 was observed in a textured polycrystalline thin film, rather than a bulk polycrystal with random grain orientations as considered above. To rationalize this case, we examine a representative through-thickness circular grain embedded in a homogeneous matrix, with the corresponding $E^{hkl}$ given in Appendix A. The relation between stiffness and orientation remains qualitatively consistent with the bulk case despite numerical differences. According to this analysis, AGG in an out-of-plane [001]-textured Ni thin film leads to a preferential in-plane [100] alignment along the loading direction, consistent with the experimental observations in Fig. 1.

**Table 1.** Single-crystal elastic constants of FCC Ni and BCC W and Cr (Zhang et al., 2020). The Zener anisotropy ratio $A_Z$ is defined as $2C_{44}/(C_{11} - C_{12})$.

| Crystal | $C_{11}$ (GPa) | $C_{12}$ (GPa) | $C_{44}$ (GPa) | $A_Z$ |
|---|---|---|---|---|
| Ni (FCC) | 246.5 | 147.3 | 124.7 | 2.51 |
| W (BCC) | 522.4 | 204.4 | 160.8 | 1.01 |
| Cr (BCC) | 339.8 | 58.6 | 99.0 | 0.70 |

## 4. Phase-field model

While the reduction of orientation-dependent elastic energy analyzed in Section 3 provides a rationale for AGG, the kinetic evolution of AGG is governed by local grain geometry and stress state, and can be effectively revealed using phase-field simulations. Following Chen et al. (Chen and Yang, 1994) and Tonks et al. (Tonks and Millett, 2011; Tonks et al., 2010), we developed a phase-field model of AGG primarily driven by a reduction in elastic energy.

In this phase-field model, the polycrystal structure is represented using a set of order parameters $(\phi_1, \phi_2, \dots, \phi_N)$, where each order parameter defines a family of grains with a specific orientation, taking the value $\phi_i = 1$ inside the $i$-th family and $\phi_i = 0$ elsewhere. The phase-field free energy incorporates the elastic contribution and is expressed as

$$F = \int_V \left[ \mu \sum_{i=1}^{N} \left( \frac{\phi_i^4}{4} - \frac{\phi_i^2}{2} \right) + \frac{3}{2}\mu \sum_{i=1}^{N} \sum_{j>i}^{N} \phi_i^2 \phi_j^2 + \frac{\mu}{4} + \frac{\beta}{2} \sum_{i=1}^{N} (\nabla \phi_i)^2 + W_{\mathrm{el}} \right] \mathrm{d}V \qquad (5)$$

where $\mu$ is the free energy coefficient and $\beta$ is the gradient energy coefficient (Chen and Yang, 1994; Tonks and Millett, 2011; Tonks et al., 2010). In Eq. (5), $W_{\mathrm{el}}$ is the elastic strain energy density

$$W_{\mathrm{el}} = \frac{1}{2} C_{ijkl} \varepsilon_{ij} \varepsilon_{kl} \tag{6}$$

where $C_{ijkl}$ is the elastic stiffness tensor, $\varepsilon_{ij}$ is the strain tensor defined as $\varepsilon_{ij} = \frac{1}{2}\left(u_{i,j} + u_{j,i}\right)$, with $u_i$ being the displacement vector. The stress tensor $\sigma_{ij}$ is given by the generalized Hooke's law, $\sigma_{ij} = C_{ijkl}\varepsilon_{kl}$, and satisfies the static equilibrium condition

$$\sigma_{ij,j} = 0 \tag{7}$$

The evolution of the order parameter $\phi_i$ follows the Allen-Cahn equation (Cahn and Allen, 1977)

$$\frac{\partial \phi_i}{\partial t} = -L \frac{\delta F}{\delta \phi_i}, \qquad i = 1, 2, \dots, N \tag{8}$$

where $L$ is a kinetic coefficient related to GB mobility (Moelans et al., 2008) and is assumed constant and orientation-independent. The nonuniform distribution of $\phi_i$ leads to spatial variation in the elastic stiffness tensor

$$C_{ijkl}(\phi_1, \phi_2, \dots, \phi_N) = \frac{\sum_{\alpha=1}^{N} h(\phi_\alpha) C_{ijkl}^{(\alpha)}}{\sum_{\alpha=1}^{N} h(\phi_\alpha)} \tag{9}$$

where $C_{ijkl}^{(\alpha)}$ is the elastic stiffness tensor of the $\alpha$-th grain family expressed in the global basis. The interpolation function $h(\phi_\alpha)$ satisfies $h(1) = 1$ and $h(0) = 0$, with vanishing derivatives at both ends,

$$h(\phi_\alpha) = \frac{1}{2}\left[1 + \sin\left(\left(\phi_\alpha - \frac{1}{2}\right)\pi\right)\right] \tag{10}$$

This phase-field model is applied to a Ni thin film with a [001] out-of-plane texture and random in-plane grain orientations (Fig. 3). The film is subjected to uniform, displacement-controlled loading along the $x$-direction, with a constant tensile strain $\varepsilon_x$ prescribed on the left and right boundaries, and strains in the $y$ and $z$ directions are unconstrained, producing a time-dependent displacement field, $u_i(x, y, z, t)$. The order parameter field $\phi_i$ satisfies homogeneous Neumann

boundary conditions

$$\nabla\phi_i \cdot \boldsymbol{n}_\mathrm{B} = 0 \quad \text{on} \quad \partial \mathrm{V}, \qquad i = 1,2,\dots,N \tag{11}$$

where $\boldsymbol{n}_\mathrm{B}$ denotes the unit normal vector to the boundary.

The simulation domain initially contains 877 through-thickness grains of nearly uniform size (~600 nm), generated by Voronoi tessellation (Miles, 1972). As shown in Fig. 3, the in-plane grain orientation is defined by rotating the [100] direction from the *x*-axis about the *z*-axis by an angle $\theta_3$, ranging from 0 to 45° due to crystal symmetry. Sixteen representative orientations, $\theta_3 \in \{0°, 3°, 6°, \dots, 45°\}$, are randomly assigned to the grains (Fig. 3). Each discrete value represents a constant orientation interval. The domain is discretized on a uniform 400 × 400 × 1 mesh of linear brick elements with an in-plane mesh size of 42.5 nm, and the time step $\Delta t$ is set to 4.5 s. The phase-field model parameters are listed in Table 2. These values correspond to a specific GB energy of $\gamma_\mathrm{GB} = 1.0\ \mathrm{J/m^2}$, a characteristic length of $l_\mathrm{GB} = 50\ \mathrm{nm}$, and a GB mobility of $M = 5 \times 10^{-3}\ \mathrm{\mu m/(s \cdot MPa)}$. Following (Moelans et al., 2008), the model parameters are related to physical parameters through $L = M\gamma_\mathrm{GB}\beta^{-1}$, $\mu = \frac{3}{\sqrt{2}}\gamma_\mathrm{GB} l_\mathrm{GB}^{-1}$ and $\beta = \frac{3}{\sqrt{2}}\gamma_\mathrm{GB} l_\mathrm{GB}$. Note that the nominal interfacial length $l_\mathrm{GB}$ is a model parameter rather than the actual GB width. Although the energy of general GBs in Ni typically varies between 0.8 and 1.4 $\mathrm{J/m^2}$ (Rohrer et al., 2010) and depends on the local GB orientation and inclination (Gao et al., 2026), we adopt a constant value to simplify the model. The GB mobility is estimated from post-mortem experimental observations and is treated as an effective mobility. While this choice accelerates simulated kinetics to a computationally accessible timescale, it does not alter the thermodynamic energy landscape and thus does not affect the predicted orientation selection trend.

Eqs. (7) and (8) are solved using a staggered scheme. At the start of each time step, the displacement field $u_i$ is computed to satisfy the static equilibrium condition, Eq. (7). The resulting displacement is then used to update the order parameter $\phi_i$ via the Allen-Cahn equation, Eq. (8). The weak forms of the governing equations and their finite element discretization are implemented using the open-source package FEniCS (Logg et al., 2012) (see Appendix B for details). Details of the mesh and time-step convergence analyses are provided in Supplementary Section 2.1. Both $u_i$ and $\phi_i$ are discretized using linear elements.

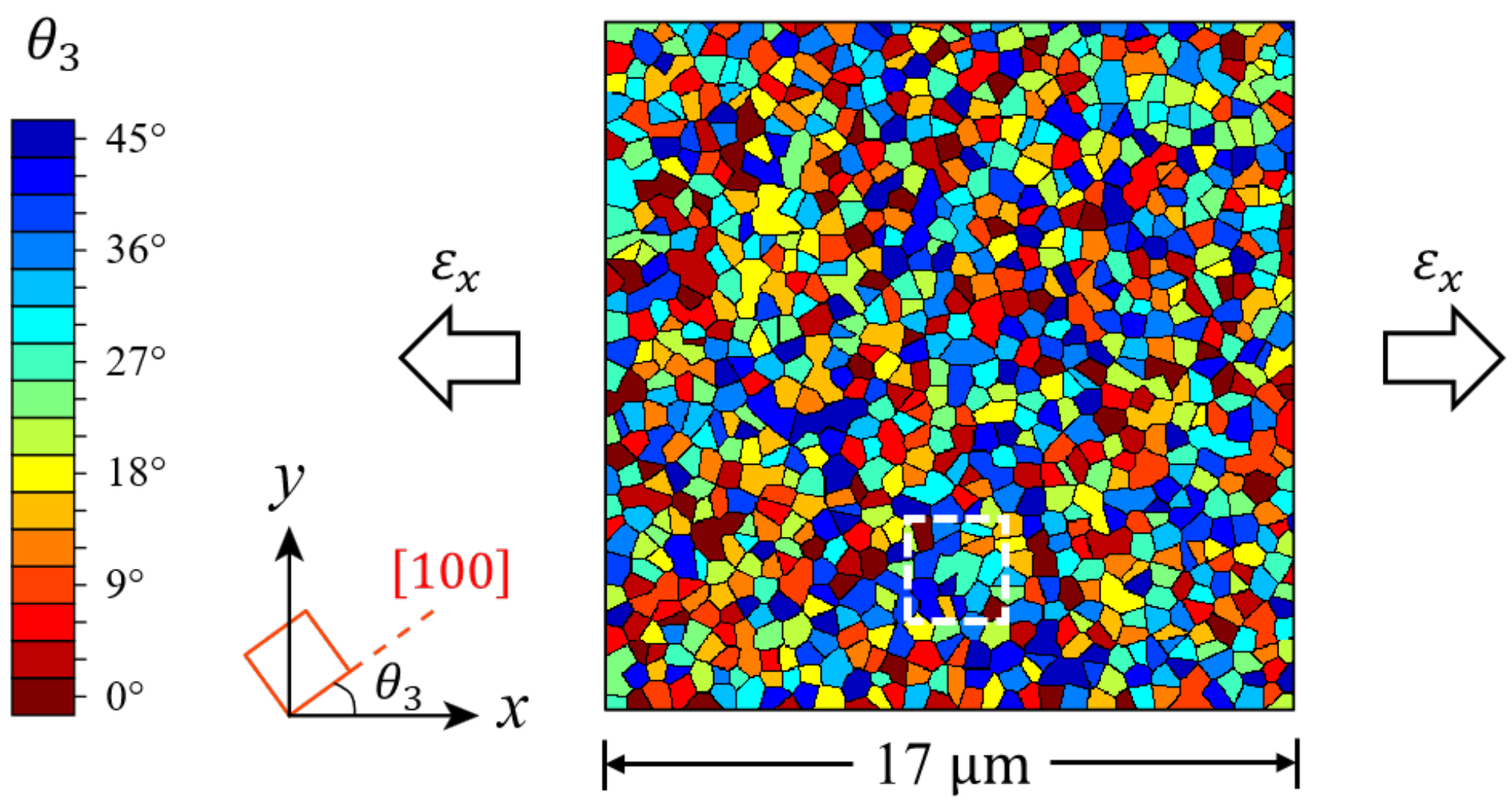

**Fig. 3.** Initial polycrystal structure with a [001] out-of-plane texture and random in-plane grain orientations. The in-plane grain orientations are described by the angle $\theta_3$ between the grain's [100] direction and the loading direction along the *x*-axis. Grains with $\theta_3 = 0°$ have their [100] aligned with the *x*-axis, while those with $\theta_3 = 45°$ have their [110] direction aligned with the *x*-axis. A prescribed displacement is applied on the right boundary to control the axial strain $\varepsilon_x$.

**Table 2.** Phase-field model parameters.

| $\mu$ | $\beta$ | $L$ |
|---|---|---|
| 42 MPa | $0.1\ \mathrm{MPa \cdot \mu m^2}$ | $0.05\ (\mathrm{s \cdot MPa})^{-1}$ |

## 5. Results and discussion

### 5.1. AGG in Ni

The initial microstructure was first relaxed for 450 s at zero applied strain and remained essentially unchanged, with no appreciable grain growth or preferential texture. Fig. 4 shows the subsequent simulation results of AGG in a strained Ni thin film with an initial [001] out-of-plane texture and random in-plane grain orientations. At $t = 0$, the grain structure shows no preferential in-plane orientation (Fig. 4(a), identical to Fig. 3). When a constant tensile strain of $\varepsilon_x$= 0.25% is applied, preferential grain growth emerges within 450 s, with red grains ($\theta_3 \approx 0°$) expanding rapidly (Figs. 4(b-c)). This trend is further supported by the evolution of the areal fraction of grain orientations. By 2025 s, a dominant in-plane [100] texture aligned with the loading direction is fully developed (Fig. 4(d)). Since the experimental characterization of AGG is post-mortem, the comparison with simulations is focused on microstructural evolution trends. The simulation results show the preferential growth of grains with elastically compliant orientations and the development of a loading-aligned in-plane [100] texture. They agree with the experimental observations in Section 2 and the system-level elastic energy analysis in Section 3.

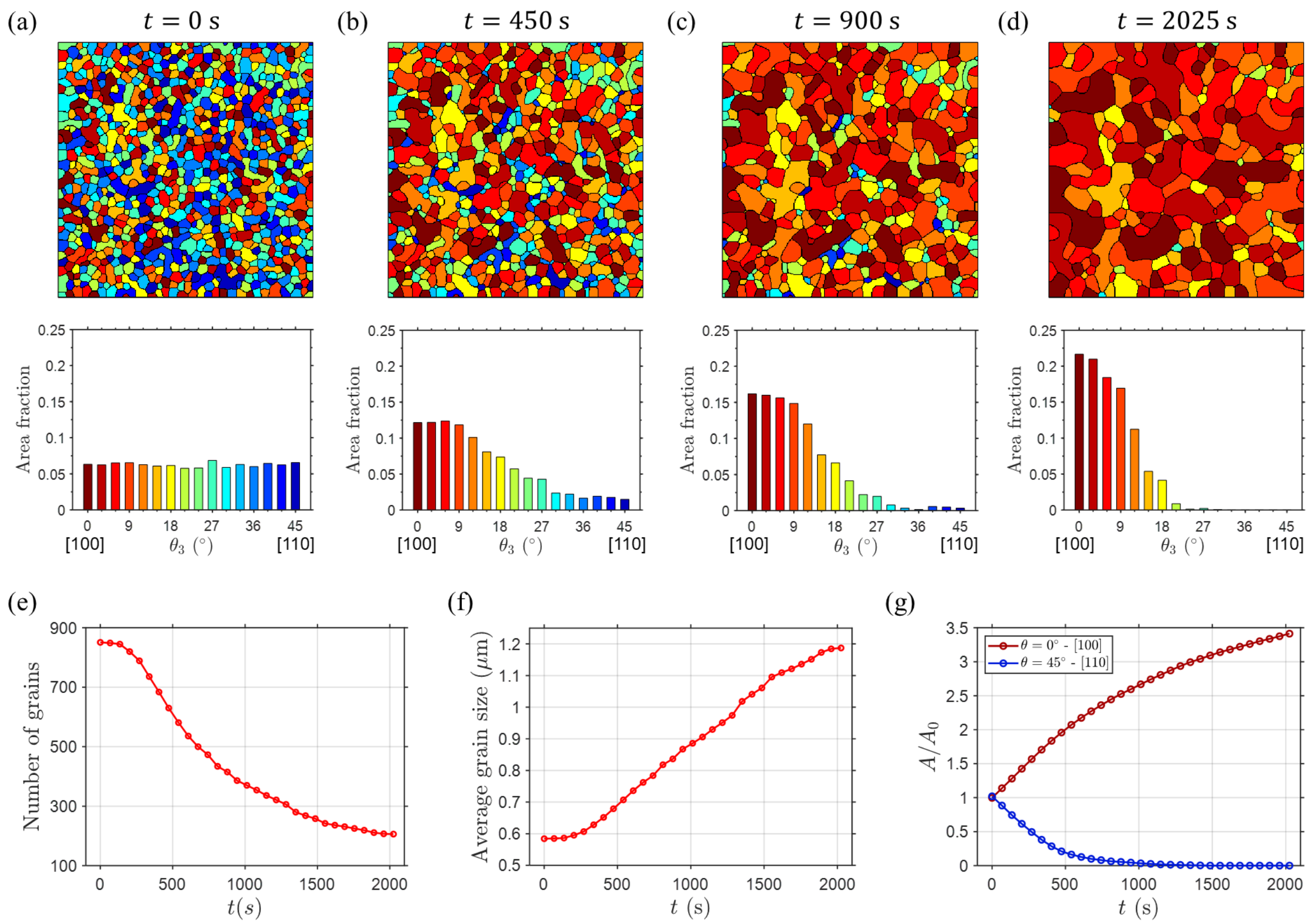

**Fig. 4.** Phase-field simulation of AGG in a polycrystalline Ni thin film. The first row shows in-plane grain orientations ($\theta_3$), using the same color scale as in Fig. 3, and the second row shows the corresponding area fraction of each grain family. (a) Initial grain structure and area distribution at $t = 0$ s, illustrating the nearly uniform in-plane orientation distribution. (b–d) Grain structures and orientation distributions at (b) $t = 450$ s, (c) $t = 900$ s, and (d) $t = 2025$ s. The increasing area fraction of red grains and the disappearance of blue grains indicate the preferential growth of [100]-oriented grains ($\theta_3 \approx 0°$), accompanied by the consumption of grains near [110] ($\theta_3 \approx 45°$). (e-g) Grain statistics of the simulation results: temporal evolution of (e) grain number, (f) average grain size, and (g) the area fractions of [100] and [110] grain families.

Consistent with the direct observations in Figs. 4(a–d), statistical analysis of the evolving microstructure supports the same conclusion. Figs. 4(e) and (f) show the temporal evolution of the total grain count and average grain size, respectively, while Fig. 4(g) tracks the normalized area fractions of the [100] and [110] grain families. During progressive coarsening, the total number of grains decreases, while the average grain size increases monotonically. However, the orientation-specific area fractions reveal a pronounced divergence: the [100] family increases steadily, whereas the [110] family decreases continuously toward zero. This asymmetric evolution demonstrates that the coarsening process is not a uniform scaling of the initial distribution. Instead, it is driven by the selective growth of elastically favorable orientations, a defining characteristic of AGG. Additional simulations based on a different initial grain structure are provided in Supplementary Section 2.2 and show the same overall trend, further confirming the consistency of the results.

### 5.2 Local GB migration

In addition to the overall AGG process, grain evolution is reflected in the local behavior of GB migration. Fig. 5 illustrates the dynamic response of a representative grain (grain #1, initial $\theta_3 = 27°$) in the Ni thin film, located in the dashed box region of Fig. 3. Its initial geometry at $t = 0$ is shown in Fig. 5(a). The evolution of its normalized area $A/A_0$ (with $A_0$ as the initial area) is plotted in Fig. 5(b). After 450 s of loading, grain #1 undergoes shape changes due to migration of its surrounding GBs. In Fig. 5(c), black lines show the current GB positions, while gray and red lines mark the initial positions. The results indicate that a GB migrates inward toward the center of grain #1 when the neighboring grain across this GB has a smaller $\theta_3$ ($< 27°$), and outward otherwise. For example, the GB between grains #1 and #3 migrates inward because grain #3 has a much smaller $\theta_3$ (12°), whereas the GB between grains #1 and #4 migrates outward due to grain #4 having a higher $\theta_3$ (39°). The relationship between GB migration direction and grain orientation $\theta_3$ will be further examined in terms of the local driving force for GB migration in Section 5.3.

The dependence of grain growth on neighboring orientations leads to a non-monotonic evolution of grain size (area) in the Ni thin film. Grain #1 initially expands but begins to shrink after ~500 s (Fig. 5(b)), driven by gradual changes in the orientations of its surrounding grains. In the early stage, grain #1 is adjacent to grains #2, #4, and #6 (Fig. 5(c)). Because grain #1 has a smaller $\theta_3$, the GBs between them migrate outward from the center of grain #1, increasing its area. As grains #2, #4, and #6 are progressively consumed by their neighbors, grains #7 and #8 become new neighbors of grain #1 (compare Figs. 5(a) and (d)). Since grains #7 and #8 have smaller $\theta_3$ values than grain #1, the GBs between them migrate inward (Fig. 5(e)), reducing the area of grain #1 (Fig. 5(b)). This non-monotonic evolution underscores that grain growth is not an isolated process of individual grains but a collective phenomenon governed by the dynamic interactions among all grains in the system.

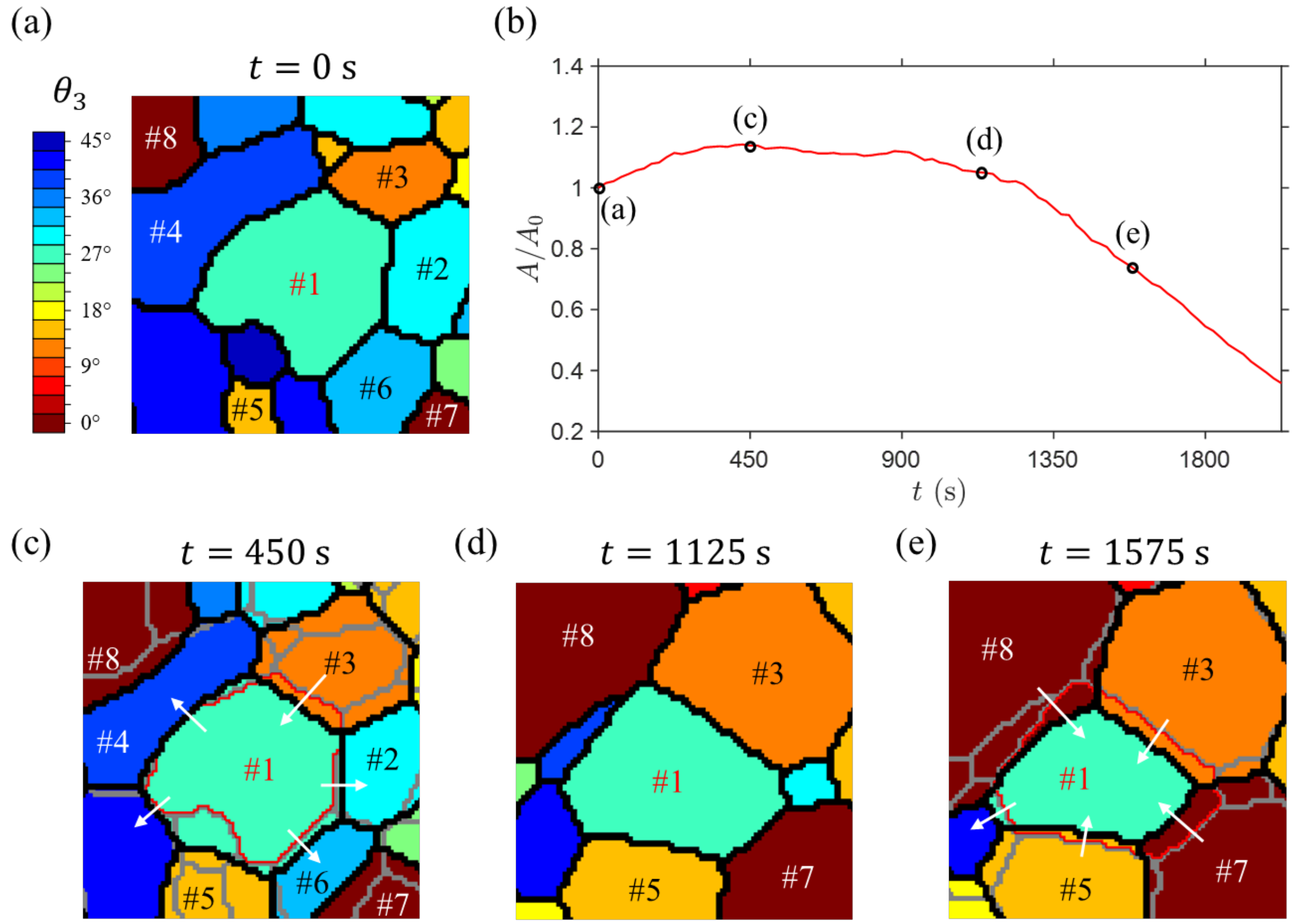

**Fig. 5.** Analysis of local GB migration for the phase-field simulation in Fig. 4. (a) Enlarged view of the region highlighted by the white dashed box in Fig. 3 at $t = 0$ s, with representative grains (#1-8) marked for analysis. (b) Time evolution of the normalized grain area $A/A_0$ of grain #1. (c) Snapshot at $t = 450$ s, showing GB migration. Black lines indicate the current GB positions, while gray and red lines mark the initial positions. White arrows denote GB migration directions. (d-e) Subsequent GB migration at (d) $t = 1125$ s and (e) $t = 1575$ s. In (e), the red and gray

lines mark the GB positions at an earlier time $t = 1125$ s. Compared to (c), the GB migration direction has reversed, correlating with orientation changes of adjacent grains and highlighting the local nature of GB migration.

### 5.3 Thermodynamic driving force for GB migration

Both the elastic and GB energy contributions to the total free energy decrease as grain growth progresses (Fig. 6(a)). Dynamic relaxation of the polycrystal during AGG drives a substantial reduction in elastic energy (green), whereas the corresponding reduction of GB energy (orange) is comparatively minor. To quantify their relative thermodynamic contributions, the reductions in elastic energy and GB energy are plotted as functions of total GB area reduction (Fig. 6(b)). The slopes of these two curves provide a measure of the corresponding global driving forces. Throughout the growth process, the driving force from a reduction in elastic energy remains consistently higher than that from GB energy, though both remain within the same order of magnitude. This observation supports the elastic energy reduction analysis in Section 3 and confirms that while both elastic and GB energies drive grain growth, the elastic contribution governs the orientation selectivity of AGG. Hence, grain orientations redistribute to lower the system's elastic energy by reducing the effective stiffness of the polycrystal along the loading direction. Meanwhile GB energy reduction facilitates general grain coarsening without imparting orientation selectivity. Because the [100] grains are the most elastically compliant in FCC Ni (Fig. 2(c)), the polycrystal naturally develops an in-plane [100] texture along the loading axis.

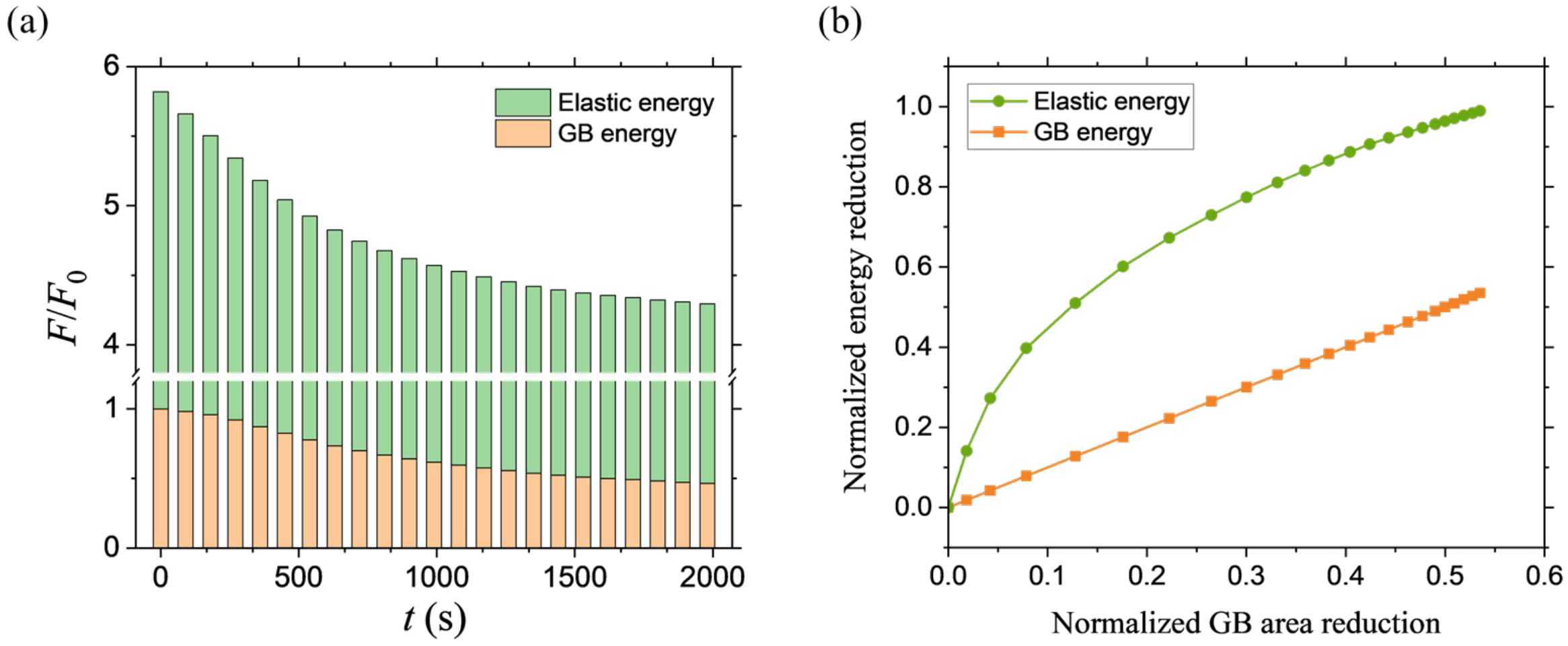

**Fig. 6.** Global driving forces of AGG. (a) Evolution of the phase-field free energy $F$, normalized

by its initial value $F_0$ before loading. The total energy $F$ is decomposed into elastic energy (green) and GB energy (orange). The $y$-axis is truncated to highlight relative changes over time. (b) Reductions in elastic energy (green) and GB energy (orange), normalized by $F_0$, as a function of the reduction in GB area, normalized by its initial value.

While a diffuse GB model is employed in phase-field simulations, its migration behavior can be interpreted through a sharp GB model by analyzing the local thermodynamic driving force and resulting migration velocity. The stress-driven contribution to GB motion is governed by the local configurational force, $\Lambda_{\mathrm{el}}$, and the GB mobility, $M$. Consistent with the linear kinetics of the phase-field model, the normal migration velocity of the boundary along its unit normal vector, $m_i$, is linearly proportional to $\Lambda_{\mathrm{el}}$

$$v = M\Lambda_{\mathrm{el}} \tag{12}$$

Here, $\Lambda_{\mathrm{el}}$ denotes the jump in the normal projection of the elastic energy momentum tensor, $P_{ij}$, across the interface along $m_i$. Following Eshelby (Eshelby, 1975), $P_{ij}$ is defined as

$$P_{ij} = W_{\mathrm{el}}\delta_{ij} - \sigma_{ik}u_{k,j} \tag{13}$$

The projection of $P_{ij}$ along $m_i$ is given by

$$\Psi = m_i P_{ij} m_j \tag{14}$$

Thus, $\Lambda_{\mathrm{el}}$ is the jump in $\Psi$ across the GB. A positive $\Lambda_{\mathrm{el}}$, and therefore a positive $v$, indicates GB migration from the side with lower $\Psi$ to the side with higher $\Psi$.

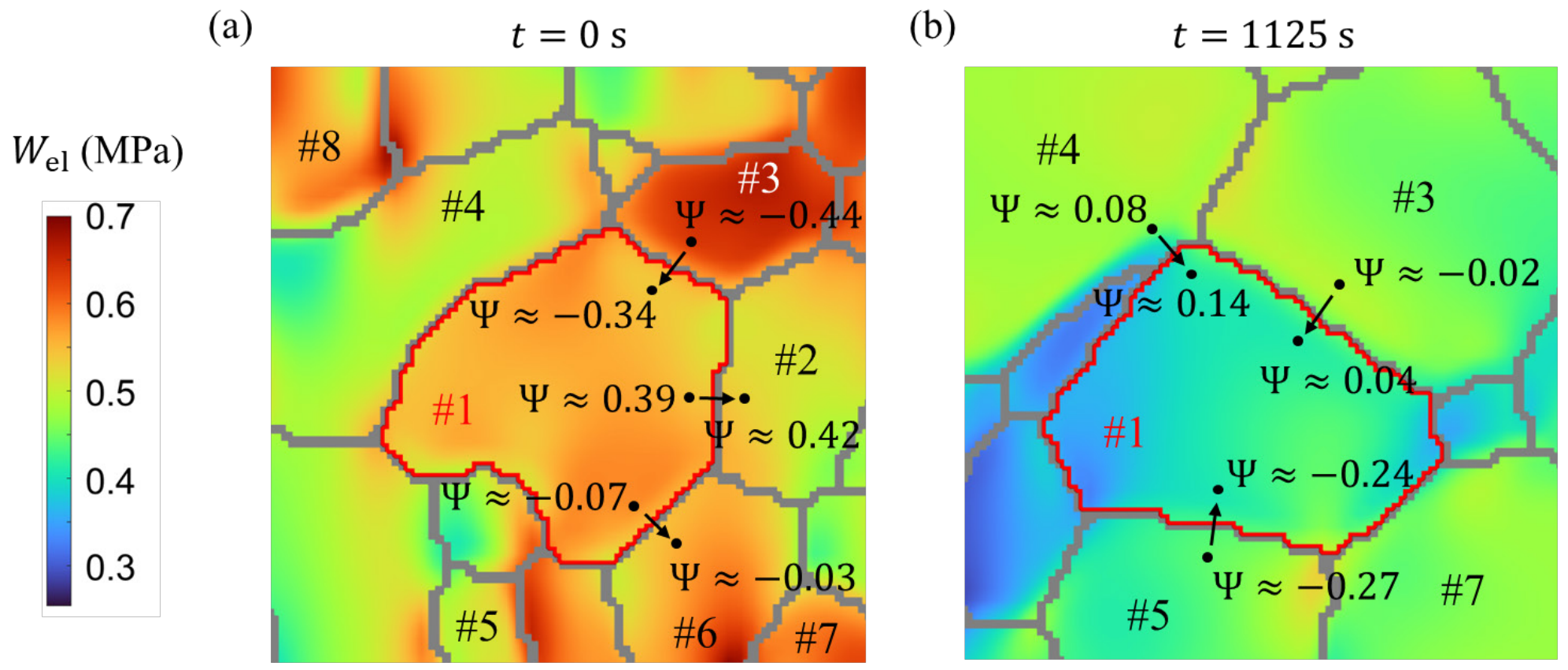

**Fig. 7.** Local thermodynamic driving force for GB migration in the Ni thin film. The contour plots

represent the spatial distribution of strain energy density at two different time steps: (a) $t = 0$ s, corresponding to the grain structure in Fig. 5(a), and (b) $t = 1125$ s, corresponding to Fig. 5(d). Grain #1 and its neighboring grains #2 to #8 are labeled for reference (same as Fig. 5(a)). Selected representative points are annotated with their respective $\Psi$ values, indicating variations in the local driving force for GB migration. The red and gray lines outline the GBs. Arrows indicate the direction of GB motion determined by comparing the values of $\Psi$ across the GB.

The local driving force for GB migration was evaluated from phase-field simulations (Fig. 7). Based on Eqs. (13) and (14), we calculated $\Psi$ at several locations near the boundary of grain #1, as shown in Fig. 7(a), which corresponds to the configuration in Fig. 5(a). In Fig. 7(a), black arrows indicate the direction of GB migration from the side of lower to higher $\Psi$, matching the migration directions shown by white arrows in Fig. 5(c). For example, the GB between grains #1 and #2 (Fig. 7(a)) moves towards grain #2, where $\Psi$ is higher, while the GB between grains #1 and #3 (Fig. 7(a)) moves towards grain #1, where $\Psi$ is higher. This trend holds in Fig. 7(b) for the moment shown in Fig. 5(d), and all results agree with the direction of $v_{\mathrm{m}}$ predicted by Eq. (12), confirming that local GB migration is driven by the local thermodynamic driving force.

Moreover, grains with [100] orientations closely aligned with the loading direction ($\theta_3$ near 0°) have lower $\Psi$ values, as can be seen by comparing grain #3 with grain #1 in Fig. 7(a) and grains #3, #4, #5 with grain # 1 in Fig. 7(b), where the orientation is shown in Figs. 5(a) and (d). These grains tend to grow, contributing to the development of an in-plane [100] texture in the simulated Ni thin film, driven by coordinated local GB movements.

Further analysis of Eqs. (12-14) reveals two driving forces of GB migration: (1) a difference in elastic energy density, $W_{\mathrm{el}}$, across the GB (from the first term in Eq. (13)), and (2) a difference in stress work (from the second term in Eq. (13)). Both play significant roles in GB migration. For example, in Fig. 7(a), the GB segment between grains #1 and #3 shows lower $W_{\mathrm{el}}$ on the grain #1 side, suggesting a driving force that resists migration toward grain #1. However, the stress work provides a stronger driving force in the opposite direction, aligning with the observed migration in Fig. 5(d) and the $\Psi$-based prediction. Hence, both driving forces are usually combined to determine the direction and velocity of GB migration. Additionally, as AGG progresses, both contributions gradually diminish when the remaining grains become increasingly similar in orientation and elastic response. This reduction in the overall thermodynamic driving force results in a significant slowdown in grain growth during the later stages of the process.

### 5.4 Role of cyclic loading in GB migration

#### 5.4.1 Time-averaged driving force

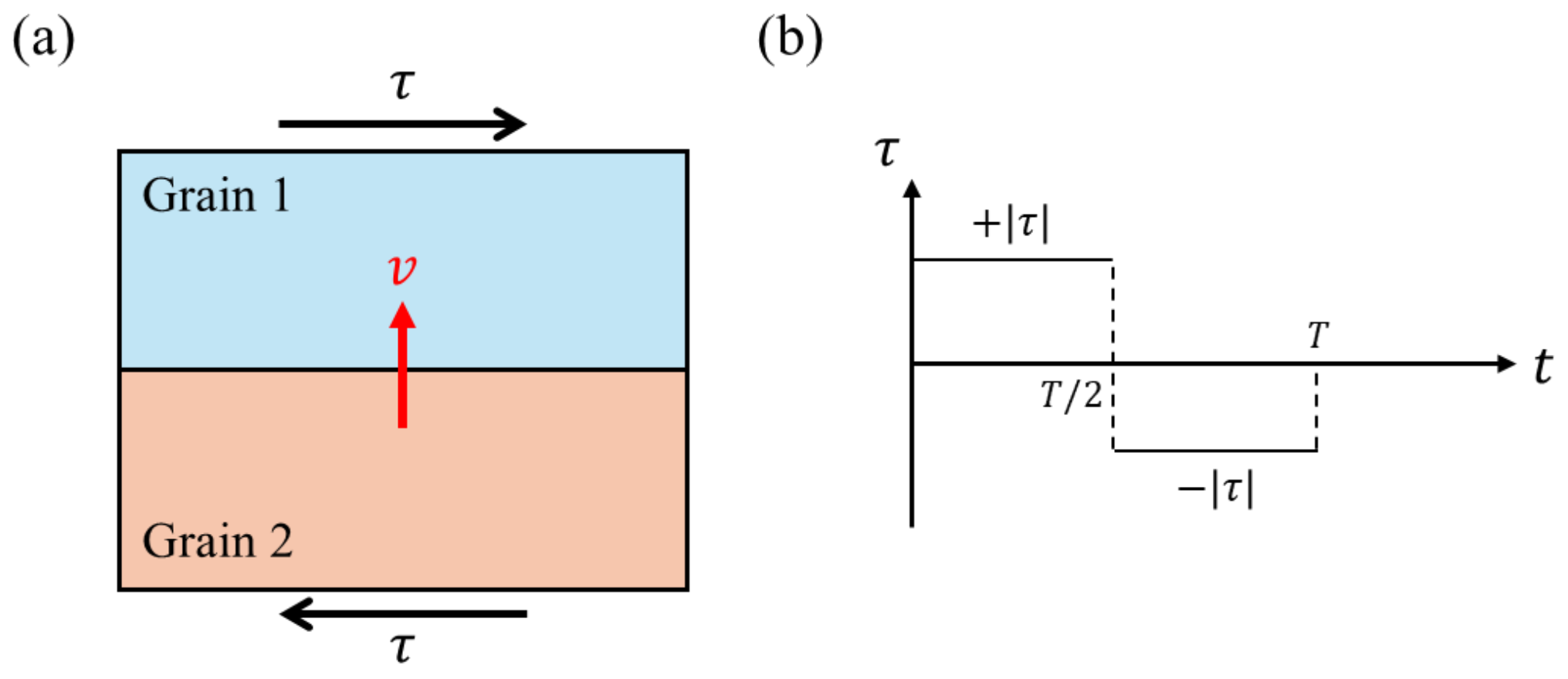

Fig. 8. Schematic illustration of GB migration under a symmetric cycle of applied shear stress ($\tau$). (a) Positive GB velocity ($v$) corresponding to GB migration from grain 2 to grain 1. (b) Applied shear stress ($\tau$) vs. time ($t$) within a loading cycle ($T$) with zero mean stress.

The mechanical driving force analyzed above accounts only for the elastic strain energy contribution in the phase-field model. Under cyclic loading, however, localized microplasticity can trigger dislocation activity, vacancy accumulation, and structural rearrangements at GBs, thereby supplying additional driving or kinetic contributions to GB migration. To clarify the roles of these processes, it is necessary to distinguish among three factors: the instantaneous driving force arising from plastic work, the cycle-averaged orientation-selective thermodynamic bias, and the defect-related mobility enhancement. Consider, for instance, GB migration from grain 2 to grain 1 under a symmetric shear stress cycle with stress magnitude $\tau$, as illustrated in Figs. 8(a-b). Let $\Delta\gamma_{\mathrm{p}}$ denote the plastic shear strain produced by the coupling of GB migration to shear deformation. The applied shear stress $\tau$ performs plastic work, giving rise to an instantaneous driving force per unit boundary area

$$\Pi_{\mathrm{pl}} = \tau\,\Delta\gamma_{\mathrm{p}} = \begin{cases} +|\tau|\,\Delta\gamma_{\mathrm{p}}, & \tau > 0 \\ -|\tau|\,\Delta\gamma_{\mathrm{p}}, & \tau < 0 \end{cases} \tag{15}$$

Meanwhile, let $\Lambda_{\mathrm{el}}$ denote the elastic-energy change associated with GB migration

$$v = M\left(\Lambda_{\mathrm{el}} + \Pi_{\mathrm{pl}}\right) = \begin{cases} M\left(\Lambda_{\mathrm{el}} + |\tau|\,\Delta\gamma_{\mathrm{p}}\right), & \tau > 0 \\ M\left(\Lambda_{\mathrm{el}} - |\tau|\,\Delta\gamma_{\mathrm{p}}\right), & \tau < 0 \end{cases} \tag{16}$$

The average GB velocity over a complete shear-loading cycle is

$$\bar{v} = \frac{1}{T}\int_0^T M\left(\Lambda_{\mathrm{el}} + \Pi_{\mathrm{pl}}\right)\mathrm{d}t = \frac{M}{T}\left(\int_0^{\frac{T}{2}}\left(\Lambda_{\mathrm{el}} + |\tau|\Delta\gamma_{\mathrm{p}}\right)\mathrm{d}t + \int_{\frac{T}{2}}^{T}\left(\Lambda_{\mathrm{el}} - |\tau|\Delta\gamma_{\mathrm{p}}\right)\mathrm{d}t\right) = M\Lambda_{\mathrm{el}} \tag{17}$$

which is equivalent to Eq. (12). In Eq. (17), it is assumed that the effective mobility $M$ and the plastic shear strain $\Delta\gamma_{\mathrm{p}}$ remain constant throughout the loading cycle. Since $\Lambda_{\mathrm{el}}$ depends quadratically on the applied stress, it does not change sign between the positive and negative half-cycles. This result for $\bar{v}$ holds true even if the instantaneous velocity temporarily becomes negative. Thus, although the plastic work term $\Pi_{\mathrm{pl}}$ acts as an instantaneous driving force during each half-cycle, its net contribution to GB migration averages to zero over a complete symmetric loading cycle. This cancellation does not imply that the plastic contribution is negligible. On the contrary, cyclic loading still drives defect storage, release, and structural rearrangements at the GB. Since these processes can lower local kinetic barriers and enhance GB mobility, our model incorporates these effects into the effective GB mobility $M$ (see Section 5.4.2 for a more detailed discussion). The analysis above can be readily extended to other symmetric loading profiles, such as triangular or sinusoidal waveforms. On this basis, we propose that the complex effects of cyclic fatigue can be effectively characterized by using an elastic loading mechanism within the framework of our phase-field model.

### 5.4.2 Mobility enhancement

As discussed above, the reduction in elastic energy under applied loading provides the thermodynamic driving force for orientation-selective AGG. However, a significant question remains: why is pronounced AGG observed experimentally under cyclic loading, while static loading of comparable magnitude results in negligible microstructural evolution?

One potential explanation lies in the kinetic accessibility of GB motion. Under static loading, the mechanical state remains largely stationary once the prescribed strain is reached; consequently, GB migration is limited by existing configurational barriers. In contrast, cyclic loading repeatedly perturbs the local stress state at the boundaries. These stress reversals may trigger localized defect activity and structural rearrangements, effectively "shaking" boundary segments out of local energy minima and facilitating the propagation of otherwise immobile segments. In this sense, cyclic loading enhances the effective mobility of GBs without altering the underlying energy landscape. Accordingly, mobility enhancement may depend on frequency, leading to different effective GB mobilities, as discussed in Supplementary Section 1.2.

Furthermore, mechanical cycling alone does not impose crystallographic selectivity, as it does not inherently distinguish between boundaries of different orientations. In the absence of elastic anisotropy, enhanced boundary motion would result in accelerated, but likely uniform, grain growth rather than AGG. The observed preferential growth of [100] grains is thus fundamentally driven by orientation-dependent elastic energy, while cyclic loading primarily affects the rate at which this energetically favored growth becomes kinetically accessible.

In our phase-field model, GB kinetics are represented by a constant mobility coefficient, $L$. While the model does not explicitly account for cycle-dependent phenomena, such as defect accumulation at boundaries, it isolates the role of elastic energy reduction in determining orientation selectivity. The simulations show that elastic anisotropy alone is sufficient to produce orientation-selective AGG. Within this framework, cyclic loading serves as the activation mechanism for boundary migration, whereas elastic anisotropy dictates the orientation selection.

### 5.5 Sensitivity of elastic anisotropy

To further examine the role of elastic anisotropy, we extended the phase-field model to polycrystals with varying elastic anisotropies, selecting the representative cubic crystals of BCC W and Cr for comparison with FCC Ni. As shown in Table 1, the anisotropy ratio $A_Z = 2C_{44}/(C_{11} - C_{12})$ differs markedly among the three metals: $A_Z = 2.51 > 1$ for Ni, $A_Z = 1.01 \approx 1$ for W, and $A_Z = 0.70 < 1$ for Cr (Zhang et al., 2020). These contrasts in $A_Z$ provide a direct means to assess the role of elastic anisotropy in AGG.

The same GB energy and mobility values as those used for Ni were adopted for W and Cr.

This was a deliberate methodological choice to isolate the specific effect of elastic anisotropy on grain growth. Using the same initial thin-film grain structure as in Fig. 4(a), phase-field simulations under identical applied loading conditions give distinct modes of grain growth (Fig. 9). For BCC W, grain orientations remain nearly random, showing no clear preference even after 2025 s (Figs. 9(a-b)). Fig. 9(c) depicts the evolution of the system energy, where the total free energy decreases only slightly. This reduction is dominated by the decrease in GB energy, while the elastic energy remains nearly unchanged throughout the simulation. This behavior is consistent with the nearly isotropic elastic response of W; since its anisotropy ratio is close to 1, the effective stiffness along the loading direction varies negligibly with orientation. Consequently, the reduction in elastic energy does not provide a significant orientation-dependent driving force, leaving grain growth to proceed via curvature-driven coarsening. In contrast, BCC Cr develops a distinct in-plane texture (Figs. 9(d-e)). Unlike FCC Ni, however, this texture is dominated by grains with [110] orientations ($\theta_3 \approx 45°$). Although the preferred orientation differs from that of Ni, the total energy reduction in Cr (Fig. 9(f)) is again primarily driven by the elastic contribution. This result indicates that elastic anisotropy provides an orientation-dependent driving force.

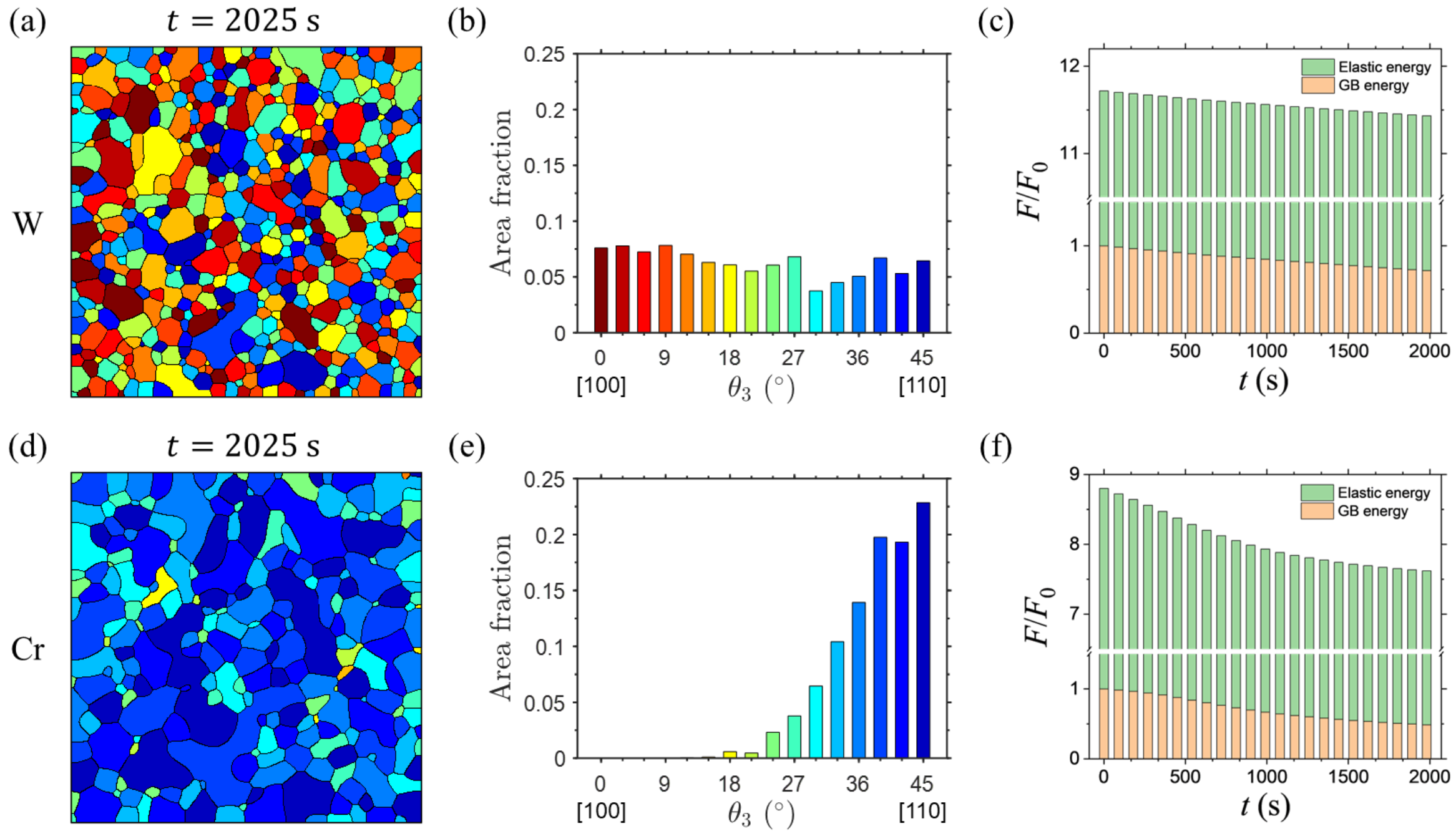

**Fig. 9.** Phase-field simulation of grain growth in polycrystalline W and Cr thin films, both initialized with the same structure as Fig. 4(a). (a–b) For W at $t = 2025$ s, larger grains grow but the in-plane orientation distribution remains nearly uniform, indicating curvature-driven coarsening without AGG. (c) Evolution of the phase-field free energy $F$ of W, normalized by its initial value $F_0$ before loading. (d–e) For Cr at $t = 2025$ s, grains oriented near [110] (i.e., $\theta_3 \approx 45°$) dominate, showing clear AGG. This contrasts with Ni thin films, where [100]-oriented grains prevail under similar loading. (f) Evolution of the phase-field free energy $F$ of Cr, normalized by its initial value $F_0$ before loading.

As discussed in Section 3, AGG lowers the elastic energy of a polycrystal under strain-controlled loading when elastically soft grains along the loading direction expand and consume stiffer ones. Fig. 10 shows the effective elastic stiffness of different grain orientations along the loading direction, $\tilde{E}(\theta_3)$ (see Appendix A for details), for Ni, W, and Cr, respectively. In Fig. 10(a), $\tilde{E}$ for Ni increases monotonically with $\theta_3$; in Fig. 10(b), $\tilde{E}$ for W is nearly orientation-independent; and in Fig. 10(c), $\tilde{E}$ for Cr decreases monotonically. Thus, in the nearly isotropic W system (Fig. 10(b)), the differences in $\tilde{E}$ between orientations are negligible, so elastic energy is not the main driving force for grain growth. Instead, grain growth is primarily driven by GB curvature, which promotes coarsening to reduce the total GB energy in the polycrystal. Because this mechanism is orientation-insensitive, it does not produce AGG with a preferred orientation. By contrast, in the strongly anisotropic Cr system, the most compliant orientation is [110] (Fig. 10(c)), leading to the development of an in-plane [110] texture, consistent with the phase-field results in Figs. 9(d-e).

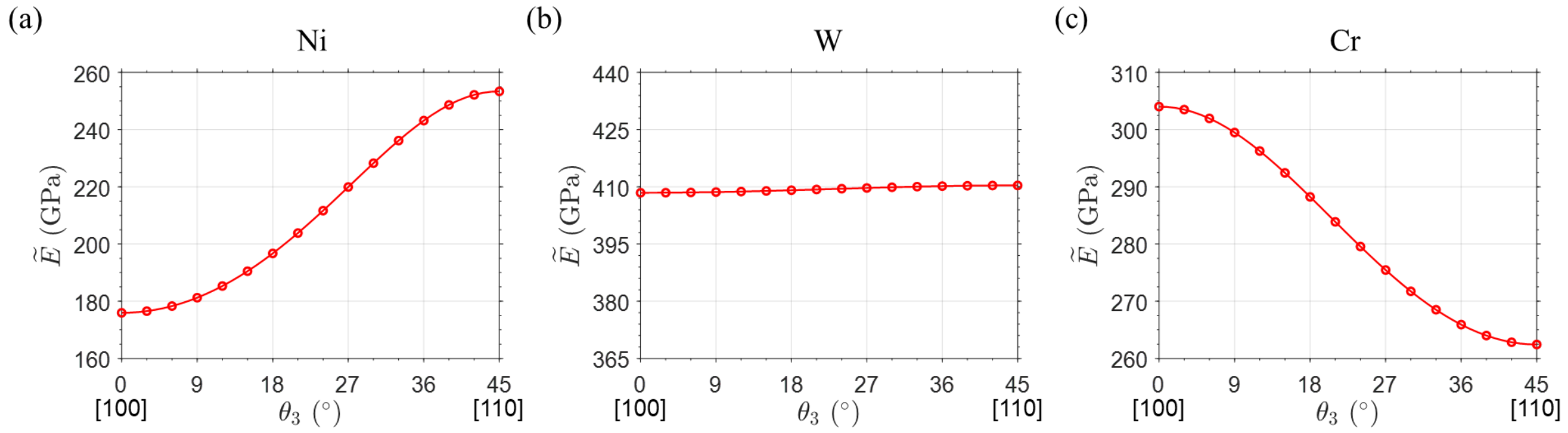

**Fig. 10**. Orientation-dependent effective elastic stiffness $\tilde{E}$ of different grain families for (a) Ni, (b) W, and (c) Cr.

We further examined the local GB migration behavior of grain #1 in polycrystalline thin films of W and Cr (Fig. 11), using the same region as in Fig. 5(a) for comparison with Ni. Phase-field simulations under identical loading conditions reveal distinct trends in grain growth. In W (Fig. 11(a)), the GBs surrounding grain #1 remain nearly unchanged. By contrast, Cr (Fig. 11(b)) shows the opposite trend to Ni: GB segments move inward when adjacent grains have larger $\theta_3$ values (near [110]) and outward when they have smaller $\theta_3$. As shown in Fig. 11(c), the area of grain #1 remains stable in W but decreases steadily in Cr, since most of its neighbors initially exhibit lower effective elastic stiffness along the loading direction. These results demonstrate that local GB migration is governed by the relative orientations and therefore the relative effective elastic stiffnesses of neighboring grains with respect to the loading axis.

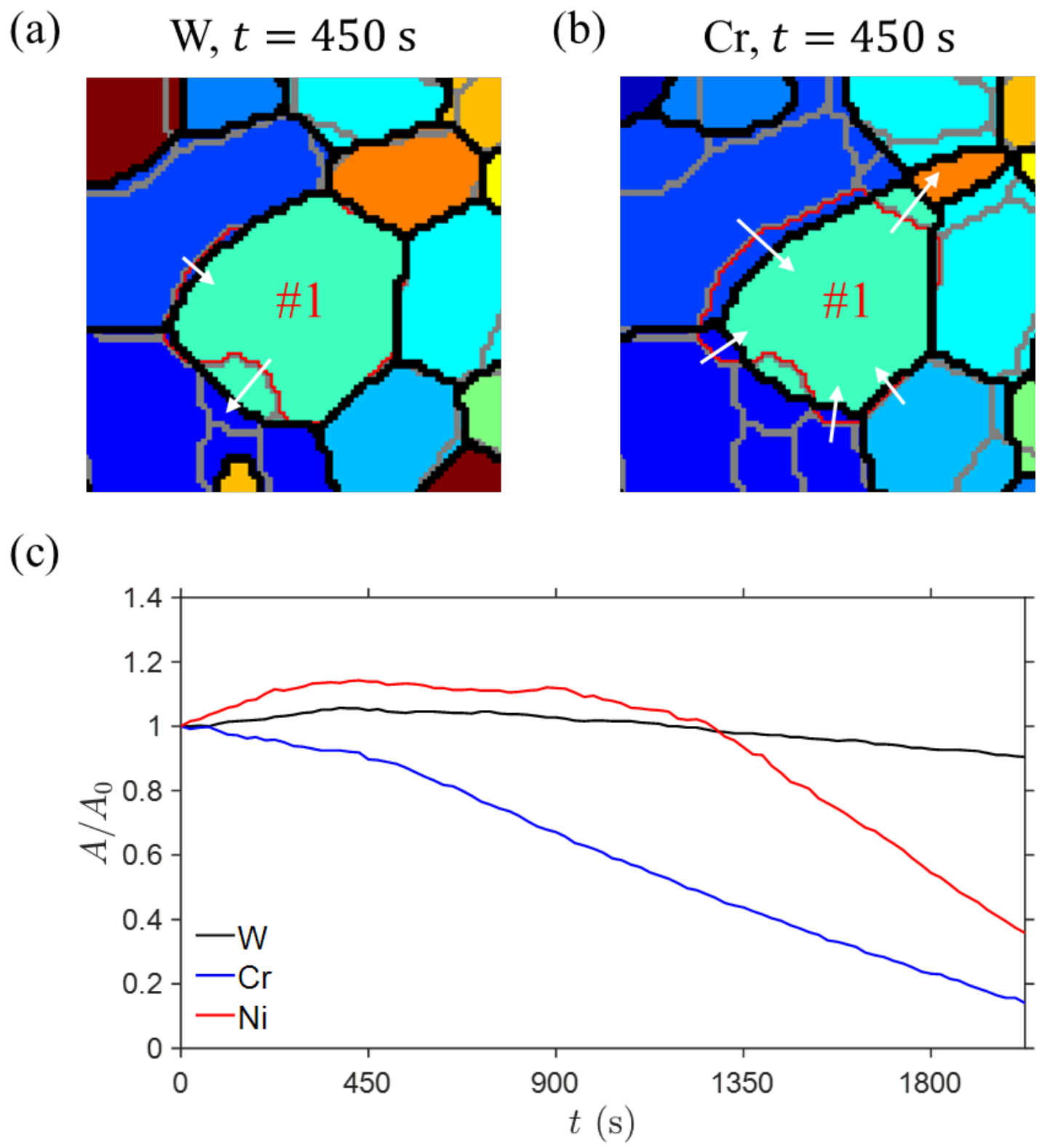

**Fig. 11.** Local GB migration at grain #1 in polycrystalline thin films of W and Cr. Snapshots at $t = 450$ s for (a) W and (b) Cr. Red lines mark the initial GBs of grain #1 at $t = 0$ s (same as Fig. 5(a)), and gray lines the initial GBs of neighboring grains. White arrows show directions of GB migration. (c) Evolution of the normalized grain area $A/A_0$ of grain #1 in W and Cr, compared with Ni.

## 6. Concluding remarks

We investigated AGG in polycrystalline thin films subjected to small-strain uniaxial loading at room temperature. Our results demonstrate that a reduction in elastic energy can provide a viable thermodynamic driving force for orientation-selective grain growth within the macroscopic elastic regime. To simulate the resulting microstructural evolution, we developed a phase-field model using the open-source FEniCS package. Phase-field simulations combined with micromechanical analysis reveal the strong influence of local grain geometry and stress state on AGG in FCC Ni thin films. We further applied this framework to BCC W and Cr, which possess distinct elastic anisotropies. The results demonstrate the broad applicability of the framework for investigating elastically driven AGG across polycrystalline systems with different crystal structures and elastic anisotropies.

This integrated approach offers several physical insights into AGG. First, a reduction in elastic energy acts as a primary thermodynamic driving force for orientation-selective AGG at room temperature without macroscopic plastic deformation. This driving force differs from mechanisms based on anisotropic GB energy or surface energy. Second, texture evolution proceeds through the cooperative migration of GBs, indicating that AGG is a collective microstructural process rather than an isolated response of individual grains. This collective mechanism explains the non-monotonic evolution of grain size. Third, the local direction of GB migration is governed primarily by the jump in the Eshelby elastic energy-momentum tensor across the interface rather than simply by the difference in elastic energy density.

These findings raise questions for future research. One key issue is the distinction between the mechanisms active under cyclic versus monotonic loading. Although the cycle-averaged thermodynamic driving force under symmetric cyclic loading can be represented by an equivalent static stress, cyclic loading modifies GB kinetics. Defect storage and release, along with structural rearrangements, may enhance the effective GB mobility, thereby promoting elastically driven AGG. Additionally, the present phase-field model does not account for free-surface relaxation, which warrants further investigation. Grain size effects also require deeper study. In ultrafine-grained Ni with grain sizes below 1 μm, the elevated yield strength expands the accessible elastic strain range, substantially increasing the thermodynamic driving force for orientation-selective AGG relative to coarse-grained counterparts.

Ultimately, investigating AGG under cyclic loading across a range of crystal structures represents an important direction for microstructure engineering. Future research should closely integrate in situ and ex situ experimental characterization with phase-field and micromechanical modeling, thereby advancing both mechanistic understanding and predictive microstructure control.

**Acknowledgements**

This work is supported by the National Science Foundation under Grant No. DMR-2224372.

**Appendix A. Effective elastic modulus of a grain family embedded in a polycrystal**

The effective elastic modulus of a grain family $\{hkl\}$, often referred to as the diffraction elastic constant, along the loading direction is defined as

$$E^{hkl} = \frac{\sigma_0}{\bar{\varepsilon}^{hkl}} \tag{A1}$$

where $\sigma_0$ is the uniaxial stress applied to the polycrystal and $\bar{\varepsilon}^{hkl}$ is the average strain of the $\{hkl\}$ grain family along the loading direction. $E^{hkl}$ can be derived analytically for a circular grain embedded in a homogenized polycrystalline matrix under plane strain deformation. By combining Eshelby's inclusion solution (Eshelby, 1957) with the self-consistent solution of the isotropic elastic moduli of a polycrystal (Qu and Cherkaoui, 2006), previous work (Zhang et al., 2020) obtained

$$\frac{1}{E^{hkl}} = U_{pqrs} n_p n_q n_r n_s \tag{A2}$$

where $n_i$ is the unit vector along the loading direction, and $U_{pqrs}$ is the constrained compliance tensor

$$U_{ijkl} = \left[I_{ijmn} + S_{ijpq}\bar{M}_{pqrs}(L_{rsmn} - \bar{L}_{rsmn})\right]^{-1}\bar{M}_{mnkl} \tag{A3}$$

with quantities defined as follows. $I_{ijmn}$ is the 4$^{\text{th}}$ order identity tensor,

$$I_{ijmn} = \frac{1}{2}\left(\delta_{im}\delta_{jn} + \delta_{in}\delta_{jm}\right) \tag{A4}$$

$S_{ijpq}$ is the Eshelby inclusion tensor for a circular inclusion, which can be found in standard micromechanics textbooks (Qu and Cherkaoui, 2006). $L_{rsmn}$ is the anisotropic stiffness tensor of the circular grain

$$L_{rsmn} = Q_{ri}Q_{sj}Q_{mk}Q_{nl}L^{(0)}_{ijkl}$$

$$L^{(0)}_{ijkl} = C_{12}\delta_{ij}\delta_{kl} + C_{44}\left(\delta_{ik}\delta_{jl} + \delta_{il}\delta_{jk}\right) + (C_{11} - C_{12} - 2C_{44})d_{ijkl} \tag{A5}$$

where $C_{11}$, $C_{12}$, and $C_{44}$ are the single-crystal elastic constants, $Q_{ij}$ is the rotation tensor mapping the local crystallographic coordinate system to the global one, $\delta_{ij}$ is the Kronecker delta, and $d_{ijkl}$ is a 4$^{\text{th}}$ order tensor with non-zero components $d_{1111} = d_{2222} = d_{3333} = 1$. $\bar{L}_{rsmn}$ and $\bar{M}_{mnkl}$ are the isotropic stiffness and compliance tensors of the homogenized polycrystal, respectively,

$$\bar{L}_{ijkl} = \frac{1}{3}(3\bar{K} - 2\bar{\mu})\delta_{ij}\delta_{kl} + \bar{\mu}\left(\delta_{ik}\delta_{jl} + \delta_{il}\delta_{jk}\right) \tag{A6}$$

$$\bar{M}_{ijkl} = \frac{1}{3}\left(\frac{1}{3\bar{K}} - \frac{1}{2\bar{\mu}}\right)\delta_{ij}\delta_{kl} + \frac{1}{4\bar{\mu}}\left(\delta_{ik}\delta_{jl} + \delta_{il}\delta_{jk}\right) \tag{A7}$$

where $\bar{K}$ and $\bar{\mu}$ are the isotropic bulk and shear moduli of the homogenized polycrystal, respectively, obtained by solving the following equations

$$\bar{K} = \frac{1}{3}(C_{11} + 2C_{12}) \tag{A8}$$

$$8\bar{\mu}^3 + (5C_{11} + 4C_{12})\bar{\mu}^2 - C_{44}(7C_{11} - 4C_{12})\bar{\mu} - C_{44}(C_{11} - C_{12})(C_{11} + 2C_{12}) = 0 \tag{A9}$$

In our phase-field model, the grain orientations are defined by rotating the [100] direction from the *x*-axis about the *z*-axis by an angle $\theta_3$, which determines the rotation tensor $Q_{ij}$. As a result, the effective elastic modulus along the loading direction $E_{hkl}$ becomes a function of $\theta_3$ and is denoted as $\tilde{E}(\theta_3)$ . A MATLAB code for calculating $\tilde{E}(\theta_3)$ is available at https://github.com/tz19/AGG2025.

## Appendix B. Numerical implementation of the phase-field model

According to the phase-field model in Section 4, combining Eqs. (5), (6) and (8) yields the governing equation for the evolution of the order parameters, $\phi_i$,

$$\frac{\partial \phi_i}{\partial t} = -L\frac{\partial f_0}{\partial \phi_i} + L\beta\nabla^2\phi_i - \frac{L}{2}\frac{\partial}{\partial \phi_i}\left(C_{pqmn}\varepsilon_{pq}\varepsilon_{mn}\right), \quad i = 1,2,\dots,N \tag{B1}$$

The variational form of Eq. (B1) is obtained by multiplying Eq. (B1) with a test function $q_i$ and integrating over the domain Ω

$$\int_\Omega \frac{\partial \phi_i}{\partial t} q_i \, dV + \int_\Omega L\frac{\partial f_0}{\partial \phi_i} q_i \, dV - \int_\Omega L\beta q_i \, \nabla^2\phi_i \, dV + \int_\Omega \frac{L}{2} q_i \frac{\partial}{\partial \phi_i}\left(C_{pqmn}\varepsilon_{pq}\varepsilon_{mn}\right) dV = 0 \tag{B2}$$

Applying integration by parts yields

$$\int_\Omega \frac{\partial \phi_i}{\partial t} q_i \, dV + \int_\Omega L\frac{\partial f_0}{\partial \phi_i} q_i \, dV + \int_\Omega L\beta \phi_{i,j} q_{i,j} \, dV + \int_\Omega \frac{L}{2} q_i \frac{\partial}{\partial \phi_i}\left(C_{pqmn}\varepsilon_{pq}\varepsilon_{mn}\right) dV = 0 \tag{B3}$$

For time discretization, we adopt the forward Euler method with time step $\Delta t$. The order parameter $\phi_i$ from the previous time step is denoted $\phi_i^{(t-\Delta t)}$, and its value at the current time step is $\phi_i^{(t)}$. The solvable weak form of Eq. (B1) becomes

$$\int_\Omega \left( \frac{\phi_i^{(t)} - \phi_i^{(t-\Delta t)}}{\Delta t} q_i + L\frac{\partial f_0}{\partial \phi_i^{(t)}} q_i + L\beta \phi_{i,j}^{(t)} q_{i,j} + \frac{L}{2} q_i \frac{\partial}{\partial \phi_i^{(t)}}\left(C_{pqmn}\varepsilon_{pq}\varepsilon_{mn}\right) \right) dV = 0 \tag{B4}$$

The displacement field, $u_i$, evolves according to Eq. (7). Its weak form, obtained by introducing a test function, $v_i$, is

$$\int_\Omega \sigma_{ij}\varepsilon_{ij}^{*} dV = 0 \tag{B5}$$

where the test strain function is $\varepsilon_{ij}^{*} = \frac{1}{2}\left(v_{i,j} + v_{j,i}\right)$. Eqs. (B4) and (B5) are solved using the open-source software FEniCS (Logg et al., 2012) with a staggered scheme. At each time step, the order parameter, $\phi_i^{(t)}$, is updated based on the displacement field from the previous step $u_i^{(t-\Delta t)}$, after which the displacement field, $u_i^{(t)}$, is recalculated using the updated order parameters, $\phi_i^{(t)}$. The conjugate gradient method is employed as the linear solver, with an incremental convergence criterion and a relative tolerance of $10^{-8}$. The FEniCS code for phase-field simulations of Ni thin films is available at https://github.com/tz19/AGG2025.

# Supplementary Material

# Phase-field modeling of elastically driven abnormal grain growth

Yazhuo Liu[a], Yin Zhang[b], Kunqing Ding[a], Yichen Yang[a], Alejandro Barrios[c], Xavier Maeder[d], Olivier Pierron[a], Xing Liu[e,*], Ting Zhu[a,*]

[a]Woodruff School of Mechanical Engineering, Georgia Institute of Technology, Atlanta, GA, 30332, USA

[b]School of Mechanics and Engineering Science, Peking University, 100871, Beijing, China

[c]Colorado School of Mines, Golden, CO, 80401, USA

[d]EMPA, Swiss Federal Laboratories for Materials Science and Technology, Laboratory for Mechanics of Materials and Nanostructures, Thun, 3602, Switzerland

[e]Department of Mechanical and Industrial Engineering, New Jersey Institute of Technology, Newark, NJ, 07102, USA

## S1 Experimental

### S1.1 Micro-resonator fatigue experimental setup

In the present work, the high-cycle fatigue response of ultrafine-grained (UFG) Ni microbeams was investigated using an experimental platform described previously (Barrios et al., 2022). This device has been used extensively over the past decade to investigate the small-scale fatigue behavior of electroplated Ni, including the nucleation (Kakandar et al., 2020) and propagation (Sadeghi-Tohidi and Pierron, 2015) of small fatigue cracks. Using electrostatic actuation (Baumert and Pierron, 2012, 2013), the microbeam can be driven at resonance (~8 kHz), resulting in fully reversed cyclic bending. To enable lower-frequency cycling, external actuation was applied using a piezoelectric actuator (Barrios et al., 2021). The microresonator die was mounted on the piezoelectric stage, and a micromanipulator was positioned adjacent to the resonator. Cycling the piezoelectric actuator at 0.5 Hz drove the resonator into contact with the micromanipulator, thereby imposing fully reversed bending of the microbeam. The micromanipulator was coupled to a strain-gauge load cell to measure the force required to displace

* Corresponding authors: xing.liu@njit.edu (X.L.), ting.zhu@me.gatech.edu (T.Z.)

the resonator, and the setup was aligned to maintain a strain amplitude between 0.25 and 1%.

The resonator rotation amplitude $\Theta$ was quantified directly from scanning electron microscope (SEM) images by tracking the displacement of a reference comb finger and converting it to a rotation angle using the measured radius of curvature of the finger. The applied strain amplitude $\varepsilon_a$ at the microbeam edge was calibrated with Finite element model reported recently (Jain et al., 2026). Because the present configuration does not permit *in situ* EBSD during resonant actuation, all microstructural characterization reported here is post mortem. Fatigue tests were periodically interrupted, with the drive turned off, to acquire data at selected cycle counts. EBSD orientation maps were acquired at an accelerating voltage of 20 kV using an EDAX/AMETEK EBSD system. Only indexed pixels with confidence index (CI) of at least 0.10 were retained for analysis.

**S1.2 Effect of frequency on grain growth**

To isolate the effect of loading frequency, post-mortem analyses were performed on tests conducted at low and high frequencies and interrupted at comparable cycle counts. The effect of frequency is evaluated by comparing Fig. S1(a) ($\varepsilon_x = 0.30\%$, $N = 6.0 \times 10^7$, 8 kHz) with Fig. S1(b) ($\varepsilon_a = 0.35\%$, $N = 1.6 \times 10^5$, 0.5 Hz). In both cases, grains with [100] orientation near the loading direction grow at the expense of others. In Fig. S1(b), the specimen fractured during cycling; therefore, only one remaining half of the microbeam is shown. Notably, reducing the loading frequency by approximately 42 times of magnitude does not alter the observed growth selectivity, as the [100] grain family still preferentially coarsens. Instead, the main effect of lowering the frequency is a pronounced reduction in the grain-growth rate. Correspondingly, the effective GB mobility decreases from ~$10^{-2}$ µm s$^{-1}$ MPa$^{-1}$ in Fig. S1(a) to ~$10^{-5}$ µm s$^{-1}$ MPa$^{-1}$ in Fig. S1(b).

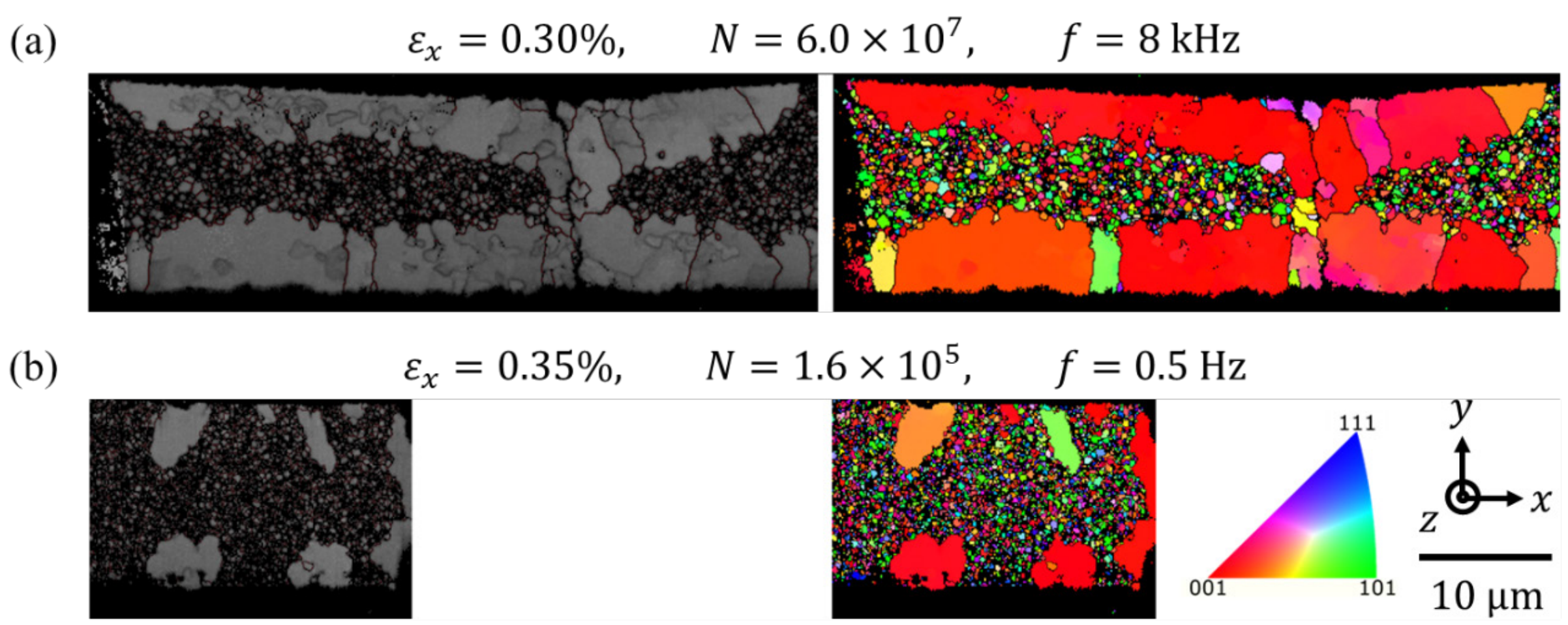

**Fig. S1.** (a-b) Band contrast images (left) and grain orientation maps along the *x* (loading) direction (right).

## S2 Simulation

### S2.1 Convergence test of phase field results

To verify the numerical convergence of the phase-field simulations, we use the strain energy, normalized by the maximum value, as a metric. We repeated the simulation of Ni with varying mesh size, $\Delta x$, and time-step size, $\Delta t$, and observed a good convergence in normalized strain energy when $\Delta x \leq 42.5$ nm and $\Delta t \leq 4.5$ s, as shown in Fig. S2. In both cases, Noticeable deviations occurred only for relatively coarse meshes or large time steps. Accordingly, the mesh and time-step sizes marked by the blue arrows were adopted in all simulations, as they provide a reasonable compromise between numerical convergence and computational efficiency.

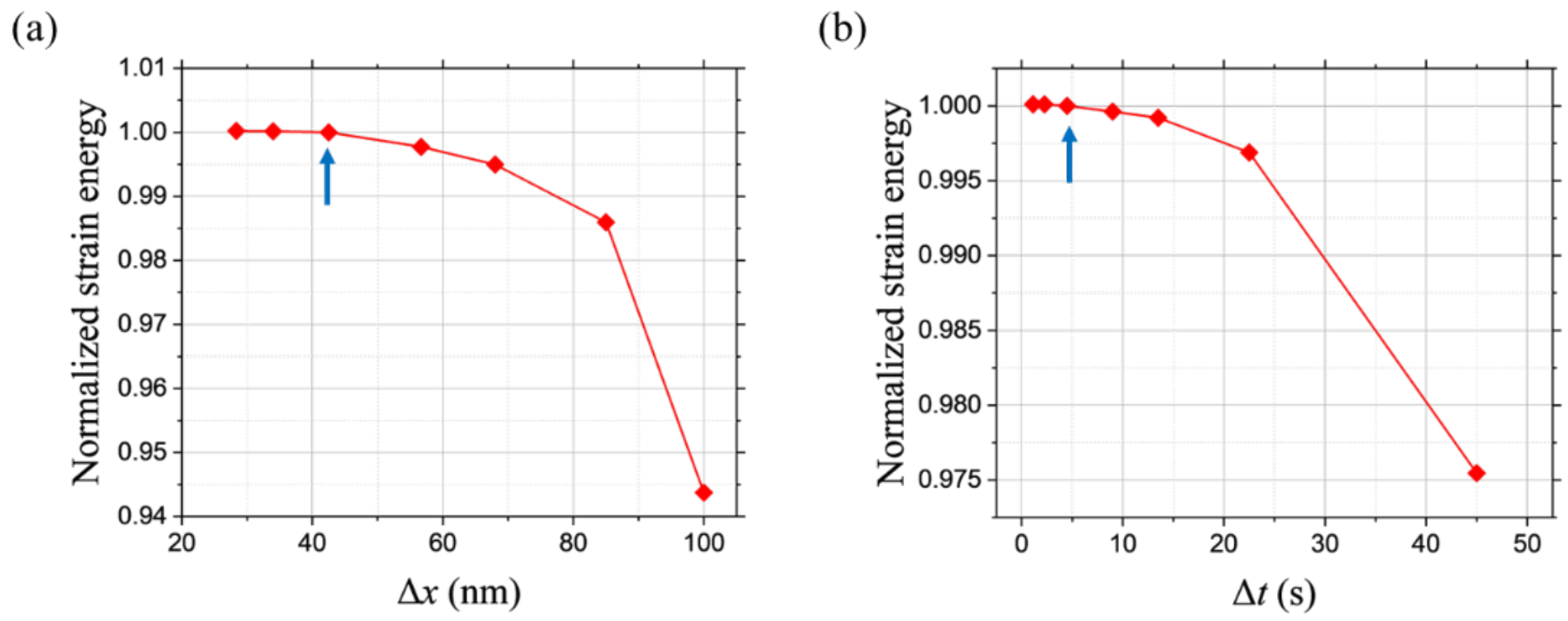

**Fig. S2.** Mesh and time-step convergence analysis. (a-b) Strain energy, normalized by the maximum value, as a function of (a) mesh size, $\Delta x$, and (b) time-step size, $\Delta t$. The adopted values are indicated by the blue arrows.

### S2.2 Effect of initial texture

An additional phase-field simulation was performed using a different initial texture (Fig. S3). Although this produced a different starting grain orientation distribution from the reference case, the same orientation-selection trend was observed. In particular, grains with [100] orientation near the loading direction remained preferentially favored during grain growth, resulting in a similar final texture. These results indicate that the observed grain selectivity is robust with respect to variations in the initial texture.

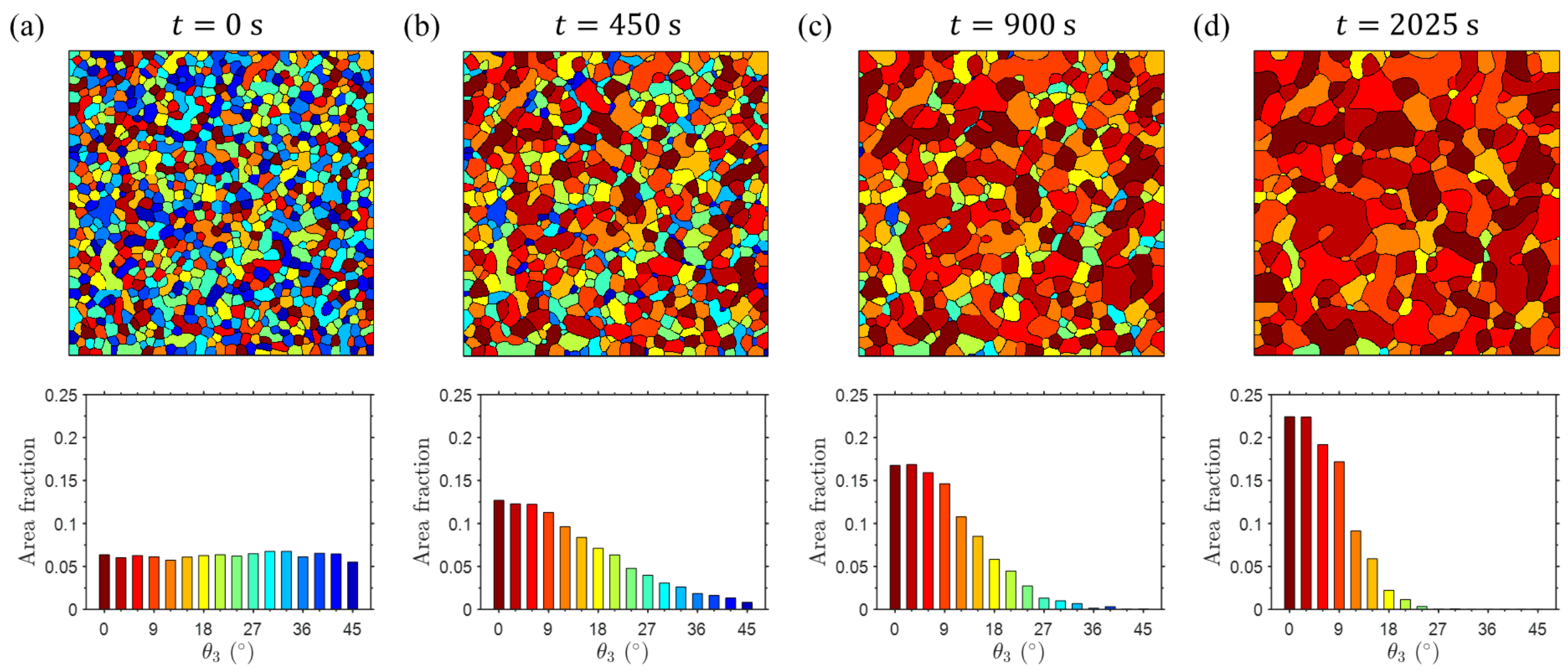

**Fig. S3** Phase-field simulation results of Ni thin film using different initial texture. The first row shows in-plane grain orientations ($\theta_3$), and the second row shows the corresponding area fraction of each grain family. These results are consistent with Fig. 4.